\documentclass[12pt]{article}
\usepackage{amsmath, amssymb,algorithm,booktabs}
\usepackage{graphicx, caption, subcaption, subfloat, float}
\usepackage[utf8]{inputenc}
\usepackage{enumerate}
\usepackage{natbib}
\usepackage{url} 

\newcommand{\blind}{0}

\DeclareMathOperator*{\argmin}{arg\,min}

\begin{document}

\bibliographystyle{natbib}

\def\spacingset#1{\renewcommand{\baselinestretch}%
{#1}\small\normalsize} \spacingset{1}


\if0\blind
{
  \title{\bf Segmented zero-inflated Poisson mixed effects model with random changepoint}
  \author{Paulo Dourado\thanks{
    Corresponding author: Paulo Dourado. Email: phdsilva@ime.usp.br }, Antonio C. Pedroso-de-Lima\hspace{.2cm}\\
    Departamento de Estat\' istica, Universidade de S\~ao Paulo, S\~ao Paulo, Brazil.\\
    and \\
    Francisco M.M. Rocha \\
    Departamento de Ciências Atuariais, Escola Paulista de Pol\' itica Economia e \\ Neg\' ocios, Universidade Federal de S\~ao Paulo, S\~ao Paulo, Brazil.}
  \date{}
  \maketitle
} \fi

\if1\blind
{
  \bigskip
  \bigskip
  \bigskip
  \begin{center}
    {\LARGE\bf Segmented zero-inflated Poisson mixed effects model with random changepoint applied to catheter-related bloodstream infections}
\end{center}
  \medskip
} \fi

\bigskip
\begin{abstract}
The COVID-19 pandemic has had a substantial impact on hospital services, as many institutions have observed a surge in  healthcare-associated infections (HAIs) despite heightened adherence to isolation protocols and hand hygiene. According to the World  Health Organization (WHO), HAIs are among the leading causes of  mortality and morbidity of hospitalized patients. This study aims to  examine the effect of the COVID-19 pandemic on the incidence of central venous catheter-related bloodstream infections (CR-BSIs) of   hospitals in the city of S\~ao Paulo. Initially we considered  segmented zero-inflated Poisson (ZIP) mixed-effects models with known changepoint, which can be estimated applying the standard framework of ZIP mixed-effects models. However, we found that the changepoint could occur at varying times across different hospitals. We present an effective iterative procedure to estimate  segmented ZIP mixed-effects models with random changepoints in a likelihood-based framework. The suggested procedure is a practical approach employing conventional computational tools for estimating standard mixed-effects zero-inflated Poisson (ZIP) models.  Prior to its implementation to the CR-BSI data, simulation studies were conducted to examine the accuracy of the estimation under various scenarios.
\end{abstract}

\noindent%
\vfill

\newpage
\spacingset{1.9} 
\section{Introduction}
\label{sec:intro}

The changepoint estimation problem has been extensively studied in the literature, and most works have focused on segmented linear (piecewise, broken-stick) regression models with one or more changepoints (Tapsoba et al., 2020). More recently, several papers have been proposed to deal with segmented models for longitudinal data in a likelihood-based framework have been proposed. We can divide the published works into two main classes: 1) segmented models based on linear-mixed effects models; and 2) segmented models based on generalized estimating equation (GEE) models (Liang and Zeger, 1986).

In the first class, we can mention, for example, the models proposed by Hall et al. (2000), who dealt with the analysis of pre-dementia phases; Jacqmin-Gadda et al. (2006), who used joint modeling to study cognitive decline and dementia risk; Lai and Albert (2014), who identified multiple changepoints in linear mixed models by integrating the expectation-maximization and dynamic programming algorithms; Muggeo et al. (2014), who presented a simple and effective iterative procedure based on Taylor expansion (linearization) to estimate segmented mixed models in a likelihood-based framework; Singer et al. (2020), who extended the model proposed by Muggeo et al. (2014) by using a smooth transition function and adding another changepoint in the linear predictor; Zhou et al. (2020), who proposed to use a mixed-effects segmented regression model and a new robust estimate to predict baseline electricity consumption in Southern California by combining the ideas of random-effects regression model, segmented regression model, and the least trimmed squares estimate; Su et al. (2021), who empirically compared segmented and stochastic linear mixed-effects models to estimate rapid disease progression in longitudinal biomarker studies.

It is important to note that the mentioned articles are based on the normal distribution. Tapsoba et al. (2020) argue that methods for estimating changepoints within the general framework of generalized linear models for correlated (or independent) data still need to be developed.

For the second class, we can mention the main works proposed by Das et al. (2016), who proposed a method using approximation by smoothing techniques for estimating multiple changepoints in linear models; Tapsoba et al. (2020), who addressed the general problem of estimating one or more unknown changepoints in the generalized linear models for independent or correlated response data. They developed methods using estimating equations to estimate the changepoints along with the other model parameters by using the derivative-free spectral algorithm proposed by La Cruz et al. (2006).

Many count data related to healthcare-associated infections (HAIs) typically exhibit zero inflation, this makes the choice of models based on the zero-inflated Poisson (ZIP) distribution a natural choice to describe the relationship between infections and covariates of interest to the study. Due to the COVID-19 pandemic, various hospital services were significantly affected, leading to regime changes in the evolution of the number of several HAIs. For this work, we consider catheter-related bloodstream infection (CR-BSI). According to P\' erez-Granda et al. (2022), information on the impact of the COVID-19 pandemic on the incidence of catheter-related bloodstream infections (CR-BSIs) is limited.

Thus, the aim of this paper is to fill some of these mentioned gaps: 1) to propose the segmented ZIP mixed-effects model with known changepoint to investigate the impact of the COVID-19 pandemic on the number of CR-BSIs; 2) we found that the changepoint could occur at different times in different hospitals. Therefore, based on the works of Muggeo et al. (2014) and Singer et al. (2020),  we present an effective iterative procedure to estimate segmented ZIP mixed-effects model with random changepoints in a likelihood-based framework. The proposed method is practical and allows the use of conventional computational tools for estimating standard ZIP mixed-effects models, such as the R package glmmTMB (Brooks et al., 2017), for example.

The rest of this paper is organized as follows. In Section \ref{sec2}, we describe the dataset that motivated the models proposed in this paper. In Section \ref{sec3}, we briefly describe the theory of the ZIP mixed-effects model. Based on this theory, we propose the segmented ZIP mixed-effects model with known changepoint to investigate the impact of the COVID-19 pandemic on the number of CR-BSIs. In Section \ref{sec4}, we propose the segmented ZIP mixed effects model with random changepoints in a likelihood-based framework with an iterative procedure for parameter estimation. In order to assess the quality of the proposed model, we conduct simulation studies. In Section \ref{sec5} we apply both models to the motivational data. Some discussion and suggestions for future work are provided in Section \ref{sec6}.

\section{Motivation} \label{sec2}

Catheter-related bloodstream infection occurs when bacteria enter the bloodstream through an intravenous catheter, and is a prevalent, serious, and costly complication of central venous catheterization. It is also the most common cause of nosocomial bacteremia (Galhot et al., 2014). Intravascular catheters are essential to modern medical practice, and they are typically inserted in critically ill patients to deliver fluids, blood products, medications, nutritional solutions, and and for hemodynamic monitoring (P\' erez-Granda et al., 2022). Central venous catheters (CVCs) are associated with a higher risk of device-related infections than any other medical device, and they are a significant cause of morbidity and mortality. They are also the leading cause of bacteremia and sepsis in hospitalized patients. In fact, most cases of catheter-associated bloodstream infections are related to the use of CVCs. In this section, we will briefly describe the dataset used in this work.

\subsection{Data description}
The dataset that will be used in this paper consists of the historical data of the number of CR-BSIs observed in large hospitals in the city of S\~ao Paulo from January 2017 to December 2020, a total of $48$ months of observation, a full description of the data can be found in Freire et al. (2023). The data were provided by researchers from the Faculty of Medicine of the University of S\~ao Paulo. The dataset consists of the following variables:
\begin{itemize}
\item $id$: Hospital identification;
\item $NAT_{i}$: Type of the $i$-th hospital: Municipal Public, Federal Public, State Public, Philanthropic, and Private;
\item $NL_{i}$: Total number of beds in the $i$-th hospital;
\item $NUTI_{i}$: Total number of ICU beds in the $i$-th hospital;
\item $BSI_{ij}$: Number of bloodstream infections in the $i$-th hospital in month $j$;
\item $CVC_{ij}$: Number of central venous catheters used in the $i$-th hospital in month $j$.
\end{itemize}
Thus, we have an unbalanced panel (repeated measure in time) with $N=76$ hospitals, so $i \in \{1,\ldots,76\}$ and $j \in \{1,\ldots,m_{i}\}$, where $m_{i}$ is the total number of observed measures for hospital $i$, totaling $3,587$ observations. The month is represented by the variable $t_{ij}$ defined in the discrete set $\{1,\ldots,48\}$, where $t_{ij}=1$ represents January 2017 and $t_{ij}=48$ represents December 2020. Furthermore, if $t_{ij}=39$, it means that we are in March 2020, where, according to the World Health Organization (WHO), on March 11, 2020, it declared that the outbreak of the new coronavirus constituted a Public Health Emergency of International Concern (PHEIC), making the beginning of the COVID-19 pandemic. 

In addition to the variables that make up the original dataset, the following variables were created for modeling and exploratory data analysis (EDA) purposes:
\begin{itemize}
\item Clustered type: $N_{i} \in \{\text{Private}, \text{Non-private}\}$, $i=1,\ldots,76$, where
\begin{eqnarray}
N_{i} = 
	\begin{cases}
	\text{Private} &\text{if}\;\;NAT_{i}\;\;\in\;\;\{\text{Private} \} \nonumber \\
	\text{Non-private} &\quad\text{otherwise}. \nonumber
	\end{cases}
\end{eqnarray} 

\item BSI rate: $R_{ij} \in (0,1)$, $i=1,\ldots,76$, $j=1,\ldots,m_{i}$ where
\begin{eqnarray}
R_{ij}=\frac{BSI_{ij}}{CVC_{ij}}.\nonumber
\end{eqnarray}
\end{itemize}

We conduct an exploratory data analysis (EDA) with the goal of identifying possible patterns that can help guide the model specification, the analysis is available in Section 1 of the Supplementary Material. Based on the EDA, we verified that: 1) The frequency distribution of the number of CR-BSIs throughout the observation period, highlighting the high frequency of zeros in the data. Therefore, the use of the ZIP model to describe the evolution of the CR-BSI rate seems reasonable. 2) We observe that there was a jump in the BSI rate shortly after the onset of the pandemic, indicating a possible regime change in the time series. This suggests that the COVID-19 pandemic generally increased the incidence of BSI in large hospitals in S\~ao Paulo. 3) The CR-BSI rate is higher in non-private hospitals. 4) The dataset for the analysis consists of $N=61$ hospitals, totaling $2,903$ observations.

\section{Segmented ZIP mixed-effects model with known changepoint} \label{sec3}

\subsection{The model} \label{sec3_1}
In a general longitudinal setting, suppose our response vector contains data from $n$ clusters such that $\mathbf{Y}^{\top}=(\mathbf{Y}_{1}, \ldots, \mathbf{Y}_{n})$, where $\mathbf{Y}_{i}^{\top}=(Y_{i1}, \ldots, Y_{im_{i}})$. Following the generalized linear mixed-effects model framework (Fitzmaurice et al., 2004), conditional on random effects $\mathbf{b}_{i}$ and $\boldsymbol{\zeta}_{i}$, the random variables $Y_{ij}$ are independent with a ZIP distribution given by Lambert (1992)
\begin{equation}
(Y_{ij}|\mathbf{b}_{i}, \boldsymbol{\zeta}_{i}) \sim 
	\begin{cases}
	0 &\quad\text{with probability}\;\; \pi_{ij} \\
	\text{Poisson}(\mu_{ij}) &\quad\text{with probability}\;\; 1-\pi_{ij}, \label{zipmixed}
	\end{cases}
\end{equation}
where $\mu_{ij} \in \mathbb{R}_{+}^{*}$ and $\pi_{ij} \in (0,1)$. In this paper, based on the EDA of the motivational data, we'll consider the following model in order to assess the effect of the COVID-19 pandemic on CR-BSI, by fixing the changepoint at March 2020 ($t_{ij}=39$) : Let $Y_{ij}$ be the count of BSI for hospital $i$ at time (month) $t_{ij}$ $(i=1,\ldots,61;\; j=1,\ldots,m_{i})$, we have that $Y_{ij}$ is represented by (\ref{zipmixed}), but the model specification for the parameters $\mu_{ij}$ and $\pi_{ij}$ are given by
\begin{align}
\label{covid1}
\begin{split}
&\text{log}(\mu_{ij})=\text{log}(O_{ij})+\beta_{0i}+\beta_{1i}t_{ij}+\delta_{i}(t_{ij}-39)_{+}+\beta_{2}N_{i}+\zeta_{t_{ij}} \\
&\text{logit}(\pi_{ij})= \gamma_{0}+\gamma_{1}N_{i}, 
\end{split} 
\end{align} 
where $O_{ij}$ is the offset which representing the CVC, $t_{ij}$ represents the time (month), $N_{i}$ represents the type of $i$-th hospital, the random effect $\zeta_{t_{ij}}$ has an interesting interpretation it captures the systemic effect of the month in all hospitals at the same time. To complete the model specification, we assume that $\beta_{0i}=\beta_{0}+b_{0i}$, $\beta_{1i}=\beta_{1}+b_{1i}$, $\delta_{i}=\delta+d_{i}$. Thus, the fixed effects of log-linear and logistic regression models are represented by the vector $\boldsymbol{\beta}^{\top}=(\beta_{0}, \beta_{1}, \beta_{2}, \delta)$ and $\boldsymbol{\gamma}^{\top}=(\gamma_{0},\gamma_{1})$. We assume that the random effects $\mathbf{b}_{i}^{\top}=(b_{0i}, b_{1i}, d_{i})$ are independent of each other and of $\boldsymbol{\zeta}_{i}^{\top}=(\zeta_{t_{i1}},\ldots,\zeta_{t_{i48}})$ for $i=1,\ldots,61$, with
\begin{eqnarray}
\mathbf{b}_{i} &\sim & \mathbf{N}_{3}(\mathbf{0}, \mathbf{G}) \nonumber \\
\zeta_{t_{ij}} &\overset{\mathrm{iid}}{\sim} & N(0,\sigma_{\zeta}^{2}). \nonumber
\end{eqnarray} 
In addition, we assume an unstructured variance-covariance matrix
\begin{eqnarray} 
\mathbf{G}=\begin{pmatrix}
\sigma_{b_{0}}^{2} & \sigma_{b_{0}b_{1}} & \sigma_{b_{0}d} \\
& \sigma_{b_{1}}^{2} & \sigma_{b_{1}d} \\
& & \sigma_{d}^{2} 
\end{pmatrix} \label{covariance1}
\end{eqnarray}  

\subsection{Estimation}
Although the model described in (\ref{covid1}) is complex, it has a parametric form, so we can use maximum likelihood estimation (MLE). The joint distribution of $\mathbf{Y}_{i}$, $\mathbf{b}_{i}$, and $\boldsymbol{\zeta}_{i}$ can be expressed as
\begin{eqnarray}
f(\mathbf{Y}_{i}, \mathbf{b}_{i},\boldsymbol{\zeta}_{i})=f(\mathbf{Y}_{i}|\mathbf{b}_{i},\boldsymbol{\zeta}_{i})f(\boldsymbol{\zeta}_{i})f(\mathbf{b}_{i}). \nonumber
\end{eqnarray}
The marginal distribution of $\mathbf{Y}_{i}$ is given by
\begin{eqnarray}
f(\mathbf{Y}_{i})=\int_{\mathbb{R}^{3+48}}f(\mathbf{Y}_{i}|\mathbf{b}_{i},\boldsymbol{\zeta}_{i})f(\boldsymbol{\zeta}_{i})f(\mathbf{b}_{i})d\mathbf{b}_{i}d\boldsymbol{\zeta}_{i}; \nonumber
\end{eqnarray}
note that 
\begin{eqnarray}
f(\mathbf{Y}_{i}|\mathbf{b}_{i},\boldsymbol{\zeta})=\prod_{j=1}^{m_{i}}f(Y_{ij}|\mathbf{b}_{i},\zeta_{t_{ij}}). \nonumber
\end{eqnarray}
Let $\boldsymbol{\theta}=(\boldsymbol{\beta}^{\top}, \boldsymbol{\gamma}^{\top}, \sigma_{b_{0}}, \sigma_{b_{1}}, \sigma_{d}, \sigma_{\zeta}, \sigma_{b_{0}b_{1}}, \sigma_{b_{0}d}, \sigma_{b_{1}d})^{\top}$, by assuming that $Y_{ij}$ are conditionally independent given the random effects $(\mathbf{b}_{i}, \zeta_{t_{ij}})$, the likelihood function for data from hospital $i$ can be written as
\begin{eqnarray}
L(\boldsymbol{\theta})=\prod_{i=1}^{61}\int_{\mathbb{R}^{3+48}}\left[\prod_{j=1}^{m_{i}}f(Y_{ij}|\mathbf{b}_{i},\zeta_{t_{ij}},\boldsymbol{\theta})\right]f(\mathbf{b}_{i}|\boldsymbol{\theta})f(\boldsymbol{\zeta}_{i}|\boldsymbol{\theta})d\mathbf{b}_{i}d\boldsymbol{\zeta}_{i}. \nonumber
\end{eqnarray}
In order to obtain the maximum likelihood estimators for the unknown parameter vector $\boldsymbol{\theta}$, we must optimize the log-likelihood function
\begin{eqnarray}
\ell(\boldsymbol{\theta})=\sum_{i=1}^{61}\ell_{i}(\boldsymbol{\theta}), \nonumber
\end{eqnarray}
where
\begin{eqnarray}
\ell_{i}(\boldsymbol{\theta})=\text{log}\left[\int_{\mathbb{R}^{3+48}}\left[\prod_{j=1}^{m_{i}}f(Y_{ij}|\mathbf{b}_{i},\zeta_{t_{ij}},\boldsymbol{\theta})\right]f(\mathbf{b}_{i}|\boldsymbol{\theta})f(\boldsymbol{\zeta}_{i}|\boldsymbol{\theta})d\mathbf{b}_{i}d\boldsymbol{\zeta}_{i}\right]. \label{loglik1}
\end{eqnarray}
Let $\hat{\boldsymbol{\theta}}$ denote the maximum likelihood estimator for $\boldsymbol{\theta}$. In this paper, the MLE is computed by maximizing the log likelihood function (\ref{loglik1}) with random effects integrated out by Laplace approximation. Model estimation is implemented using the R package glmmTMB (Brooks et al. 2017). 

\section{Segmented ZIP mixed-effects model with random changepoint} \label{sec4}
CR-BSI is complex in nature with multiple possible causes. Therefore, it is conceivable that changes in infection rates may be hospital related. We performed an exploratory analysis (not shown here) and concluded that although the introduction of COVID-19 appears to have had an important impact on CR-BSI rates, other factors may have contributed to changes in trends at hospital-specific time points. Therefore, this information served as motivation to extend the model (\ref{covid1}) to include random changepoints. At this point, we assume that the model is identifiable according to Feder's terminology (Feder, 1975).
\subsection{The model}
Suppose our response vector contains data from $n$ clusters such that $\mathbf{Y}^{\top}=(\mathbf{Y}_{1}, \ldots, \mathbf{Y}_{n})$, where $\mathbf{Y}_{i}^{\top}=(Y_{i1}, \ldots, Y_{im_{i}})$. Conditional on the random effects $\mathbf{b}_{i}$, the random variables $Y_{ij}$ are independent with ZIP distribution given by	
\begin{equation}
(Y_{ij}|\mathbf{b}_{i}) \sim 
	\begin{cases}
	0 &\quad\text{with probability}\;\; \pi_{ij} \\
	\text{Poisson}(\mu_{ij}) &\quad\text{with probability}\;\; 1-\pi_{ij}, \label{zipmixed1}
	\end{cases}
\end{equation}
where $\mu_{ij} \in \mathbb{R}_{+}^{*}$ and $\pi_{ij} \in (0,1)$. Suppose that the repeated measurements $Y_{ij}$ are associated with a time variable $t_{ij}$. We assume that all $t_{ij}$ take values from a finite set of $m$ time points $\{s_{1},\ldots,s_{m} \}$; for our motivational data, $t_{ij}$ take values from $\{1,2,\ldots,48 \}$, and experience at least one changepoint. Denote $\mathbf{X}_{ij}$ and $\mathbf{W}_{ij}$ as the design vectors for the fixed effects and $\mathbf{Z}_{ij}$ as the design vector for the random effects. The parameters $\mu_{ij}$ and $\pi_{ij}$ in (\ref{zipmixed1}) are modelled by a log-linear and logistic  regression models
\begin{align}
\label{segmodel} 
\begin{split}
&\text{log}(\mu_{ij})=\text{log}(O_{ij})+\mathbf{X}_{ij}^{\top}\boldsymbol{\beta}+\mathbf{Z}_{ij}^{\top}\mathbf{b}_{i}+\delta_{i}f(t_{ij}, \Psi_{i}(\lambda_{i})) 
\\
&\text{logit}(\pi_{ij})= \mathbf{W}_{ij}^{\top}\boldsymbol{\gamma},
\end{split}
\end{align}
where $\text{log}(O_{ij})$ is an offset and $f(.)$ is a known function representing the segmentation function; for example, Muggeo et al. (2014) consider $f(t_{ij},\Psi_{i}(\lambda_{i}))=(t_{ij}-\Psi_{i}(\lambda_{i}))_{+}$ (linear segmentation function) and Singer et al. (2020) used $f(t_{ij},\Psi_{i}(\lambda_{i}))=(t_{ij}-\Psi_{i}(\lambda_{i}))_{+}^{2}$ (quadratic segmentation function). We assume that $\delta_{i}=\delta+d_{i}$, $\lambda_{i}=\lambda+l_{i}$, and that
\begin{eqnarray}
\Psi_{i}(\lambda_{i})\equiv g(\lambda_{i}) =\frac{L_{1}+L_{2}\text{exp}(\lambda_{i})}{1+\text{exp}(\lambda_{i})}. \label{segfunction}
\end{eqnarray}
In addition, we consider that the random effects $(\mathbf{b}_{i}, d_{i}, l_{i})^{\top}$ follow a multivariate normal distribution with mean vector $\textbf{0}$ and variance-covariance matrix $\boldsymbol{\Sigma}$. The changepoint function (\ref{segfunction}) was firstly proposed by Muggeo et al. (2014), and also used by Singer et al. (2020) with aim of restricting the values of $\Psi_{i}(\lambda_{i})$ to the interval $(L_{1}, L_{2})$. Our model is an extension of the models proposed by Muggeo et al. (2014) and Singer et al. (2020) since it allows working with generalized linear and zero-inflated mixed-effects models.

\subsection{Estimation procedure}
Due to the nonlinear nature of the log-linear predictor in (\ref{segmodel}), estimation of the parameters is a difficult task, especially in a frequentist framework. Furthermore, the likelihood is not differentiable with respect to $\Psi_{i}(\lambda_{i})$ when the changepoint is random. Thus, following the ideas of Muggeo et al. (2014) and Singer et al. (2020), the segmentation function in (\ref{segmodel}) may be approximated by a first-order Taylor around $\hat{\lambda}_{i}$, that is
\begin{eqnarray}
f(t_{ij}, \Psi_{i}(\lambda_{i})) \approx f(t_{ij}, \Psi_{i}(\hat{\lambda}_{i}))+(\lambda_{i}-\hat{\lambda}_{i})\frac{\partial f(t_{ij}, \Psi_{i}(\lambda_{i}))}{\partial \Psi_{i}}\frac{\partial \Psi_{i}(\lambda_{i})}{\partial \lambda_{i}}\bigg\rvert_{\lambda_{i} = \hat{\lambda}_{i}}. \nonumber
\end{eqnarray}
with
\begin{eqnarray}
\frac{\partial f(t_{ij}, \Psi_{i}(\lambda_{i}))}{\partial \Psi_{i}} =
	\begin{cases}
	-I(t_{ij} > \Psi_{i}(\lambda_{i})) &\quad\text{linear case} \\
	-2(t_{ij}- \Psi_{i}(\lambda_{i}))I(t_{ij} > \Psi_{i}(\lambda_{i})) &\quad\text{quadratic case}, \nonumber
	\end{cases}
\end{eqnarray}
and
\begin{eqnarray}
\frac{\partial \Psi_{i}(\lambda_{i})}{\partial \lambda_{i}}=\frac{(L_{2}-L_{1})e^{\lambda_{i}}}{(1+e^{\lambda_{i}})^{2}}. \nonumber
\end{eqnarray}
Letting $U_{ij}=f(t_{ij}, \Psi_{i}(\hat{\lambda}_{i}))$ and $V_{ij}=\frac{\partial f(t_{ij}, \Psi_{i}(\lambda_{i}))}{\partial \Psi_{i}}\frac{\partial \Psi_{i}(\lambda_{i})}{\partial \lambda_{i}}\bigg\rvert_{\lambda_{i} = \hat{\lambda}_{i}}$, we may approximate the log-linear regression model (\ref{segmodel}) by
\begin{eqnarray}
\text{log}(\mu_{ij}) &\approx & \text{log}(O_{ij})+\mathbf{X}_{ij}^{\top}\boldsymbol{\beta} +\mathbf{Z}_{ij}^{\top}\mathbf{b}_{i}+\delta_{i}U_{ij}+\delta_{i}(\lambda_{i}-\hat{\lambda}_{i})V_{ij}. \label{seg_pois1}
\end{eqnarray}
Following Muggeo et al. (2014), assuming a known and approximate value for $\delta_{i}$, $\tilde{\delta}_{i}$, we can rewrite (\ref{seg_pois1}) as follows:
\begin{eqnarray}
\text{log}(\mu_{ij}) &\approx & \text{log}(O_{ij})+\mathbf{X}_{ij}^{\top}\boldsymbol{\beta}+\mathbf{Z}_{ij}^{\top}\mathbf{b}_{i}+\delta_{i}U_{ij}+(\lambda_{i}-\hat{\lambda}_{i})\tilde{\delta}_{i}V_{ij} \nonumber \\
&=& (\text{log}(O_{ij})-\hat{\lambda}_{i}\tilde{\delta}_{i}V_{ij})+\mathbf{X}_{ij}^{\top}\boldsymbol{\beta} +\mathbf{Z}_{ij}^{\top}\mathbf{b}_{i}+\delta_{i}U_{ij}+\lambda_{i}\tilde{\delta}_{i}V_{ij} \nonumber \\
&=& O_{ij}^{*}+\mathbf{X}_{ij}^{\top}\boldsymbol{\beta}+\mathbf{Z}_{ij}^{\top}\mathbf{b}_{i}+\delta_{i}U_{ij}+\lambda_{i}G_{ij}, \label{seg_pois2}
\end{eqnarray}
where $G_{ij}=\tilde{\delta}_{i}V_{ij}$ and $O_{ij}^{*}=\mbox{log}(O_{ij})-\hat{\lambda}_{i}G_{ij}$. Therefore, the variables $O_{ij}$, $U_{ij}$ and $G_{ij}$ can be considered as pseudo-offset e pseudo-covariates respectively.
Taking $\lambda_{i}=\lambda+ l_{i}$ and $\delta_{i}=\delta+d_{i}$, we rewrite (\ref{seg_pois2}) in the following form
\begin{eqnarray}
\text{log}(\mu_{ij}) &\approx & O_{ij}^{*}+\mathbf{X}_{ij}^{\top}\boldsymbol{\beta}+\mathbf{Z}_{ij}^{\top}\mathbf{b}_{i}+(\delta+d_{i})U_{ij}+(\lambda + l_{i})G_{ij} \nonumber \\
&=& O_{ij}^{*}+\mathbf{X}_{ij}^{\top}\boldsymbol{\beta}+\mathbf{Z}_{ij}^{\top}\mathbf{b}_{i}+\delta U_{ij}+\lambda G_{ij}+d_{i}U_{ij}+l_{i}G_{ij} \nonumber \\
&=& O_{ij}^{*}+\mathbf{X}_{ij}^{*\top}\boldsymbol{\beta}^{*}+\mathbf{Z}_{ij}^{*\top}\mathbf{b}_{i}^{*}, \label{seg_pois3}
\end{eqnarray} 
where $\mathbf{X}_{ij}^{*}=(\mathbf{X}_{ij},U_{ij},G_{ij})^{\top}$, $\mathbf{Z}_{ij}^{*}=(\mathbf{Z}_{ij},U_{ij},G_{ij})$, $\boldsymbol{\beta}^{*}=(\boldsymbol{\beta},\delta,\lambda)^{\top}$ and $\mathbf{b}_{i}^{*}=(\mathbf{b}_{i},d_{i},l_{i})^{\top}$. In this framework, the parameters $\mu_{ij}$ and $\pi_{ij}$ in (\ref{zipmixed1}) are modelled by a log-linear and a logistic  regression models
\begin{eqnarray}
\text{log}(\mu_{ij})&=&O_{ij}^{*}+\mathbf{X}_{ij}^{*\top}\boldsymbol{\beta}^{*}+\mathbf{Z}_{ij}^{*\top}\mathbf{b}_{i}^{*}  \label{segmodel2} \\
\text{logit}(\pi_{ij})&=& \mathbf{W}_{ij}^{\top}\boldsymbol{\gamma}. \nonumber
\end{eqnarray} 
We call this model a working ZIP mixed-effects model. It's interesting to note that this method suggests the following iterative algorithm to fit the parameters of (\ref{segmodel}).

\begin{description}
\item \textbf{Initialization:} Choose initial values for $\Psi_{i}^{(0)}=\Psi^{(0)}$, with $i=1, \ldots, N$, and compute the pseudo-covariate $U_{ij}^{(0)}=f(t_{ij}, \Psi_{i}^{(0)})$. Fit the working ZIP mixed-effects model with linear predictors
\begin{eqnarray}
\mbox{log}(\mu_{ij})&=&\mbox{log}(O_{ij})+\mathbf{X}_{ij}^{\top}\boldsymbol{\beta} +\mathbf{Z}_{ij}^{\top}\mathbf{b}_{i}+\delta_{i}U_{ij}^{(0)} \nonumber \\
\text{logit}(\pi_{ij})&=& \mathbf{W}_{ij}^{\top}\boldsymbol{\gamma}, \nonumber
\end{eqnarray}
to get preliminary estimates: $\boldsymbol{\beta}^{(0)}$, $\mathbf{b}_{i}^{(0)}$, $\boldsymbol{\gamma}^{(0)}$, $\delta_{i}^{(0)}$ and $\lambda_{i}^{(0)}=g^{-1}(\Psi_{i}^{(0)})$.
\item \textbf{Step 1:} Let $r=1$. Compute
\begin{eqnarray}
\Psi_{i}^{(r)}&=& g(\lambda_{i}^{(r-1)}), \nonumber \\
U_{ij}^{(r)}&=&f(t_{ij}, \Psi_{i}^{(r)}), \nonumber \\
V_{ij}^{(r)}&=&\frac{\partial f(t_{ij}, \Psi_{i}^{(r)}(\lambda_{i}))}{\partial \Psi_{i}^{(r)}}\frac{\partial \Psi_{i}^{(r)}(\lambda_{i})}{\partial \lambda_{i}}\bigg\rvert_{\lambda_{i} = \lambda_{i}^{(r-1)}}, \nonumber \\
G_{ij}^{(r)}&=&\delta_{i}^{(r-1)}V_{ij}^{(r)} \nonumber \\
O_{ij}^{*(r)}&=&\mbox{log}(O_{ij})-\lambda_{i}^{(r-1)}G_{ij}^{(r)}. \nonumber
\end{eqnarray}
\item \textbf{Step 2:} Fit the working ZIP mixed-effects model with linear predictors
\begin{eqnarray}
\mbox{log}(\mu_{ij})&=&O_{ij}^{*(r)}+\mathbf{X}_{ij}^{\top}\boldsymbol{\beta} +\mathbf{Z}_{ij}^{\top}\mathbf{b}_{i}+\delta_{i}U_{ij}^{(r)}+\lambda_{i}G_{ij}^{(r)} \nonumber \\
\text{logit}(\pi_{ij})&=& \mathbf{W}_{ij}^{\top}\boldsymbol{\gamma}, \nonumber
\end{eqnarray}
to get current estimates: $\boldsymbol{\beta}^{*(r)}$, $\mathbf{b}_{i}^{*(r)}$ and $\Psi_{i}^{(r)}=g(\lambda_{i}^{(r)})$.
\item \textbf{Step 3:} Let $r=r+1$. Repeat steps 1 and 2, until the relative difference between the log likelihoods of the last two fits is small ($\epsilon<10^{-5}$).
\end{description}
The proposed iterative procedure makes maximization quite feasible, as the standard ZIP mixed-effects model has to be iteratively fitted, in this paper we use the R package glmmTMB (Brooks et al. 2017). However, in practice and in simulation studies, we found that the quality of the fitted values depends on the initial values $\Psi^{(0)}$; in some cases, it does not converge and in other cases, it can lead to local maxima (Muggeo et al. 2014; Jacqmin-Gadda et al. 2006; Singer et al. 2020). To circumvent this problem, we applied the grid search method (Lerman, 1980; Singer et al., 2020), which is commonly used for hyperparameter tuning in machine learning models, to select the initial value for $\Psi^{(0)}$. Roughly speaking, let $S=\{\psi_{1}=L_{1},\psi_{2},\ldots,\psi_{m-1},\psi_{m}=L_{2} \}$, with $\psi_{1}<\ldots<\psi_{m}$ and $\psi_{i} \in [L_{1},L_{2}]$ known, for $i=1,\ldots,m$, a finite set of $m$ equally spaced points representing the grid of values for $\Psi^{(0)}$. The idea of this method consists in systematically evaluate the mean squared error (MSE) between observed and predicted individual values for each $\{\psi_{i}:i=1,\dots,m \}$ and, for algorithm initialization, we choose $\Psi^{(0)}$ as 
\begin{eqnarray}
\Psi^{(0)}= \argmin_{\Psi \in S}\mathbb{E}_{\Psi}\left[(y_{ij}-\hat{y}_{ij})^{2}\right], \nonumber
\end{eqnarray}
where $\hat{y}_{ij}=\mathbb{E}(Y_{ij}|\mathbf{b}_{i}, \boldsymbol{\zeta})=(1-\hat{\pi}_{ij})\hat{\mu}_{ij}$. When $m$ grows the algorithm tends to escape local maxima of log-likelihood at the expense of computational complexity. The algorithm was implemented in R and the code is available in the Supplementary Material.

\subsection{Simulation study}
A simulation study was carried out to evaluate the performance of the proposed framework with respect to the point estimation of the model parameters, especially the known effect $\lambda$ related to the changepoint, in two different scenarios.

\textbf{Scenario 1:} The data were generated from the segmented ZIP mixed-effects model (\ref{zipmixed1}). The parameters were simulated by the following linear predictors
\begin{eqnarray}
\text{log}(\mu_{ij}) &=& \beta_{0i}+\beta_{1i}x_{ij}+\delta_{i}(x_{ij}-\Psi_{i})_{+} \nonumber \\
\text{logit}(\pi_{ij}) &=& \gamma_{0}, \nonumber
\end{eqnarray}  
where $i=1,\ldots,2000$, $j=1,\ldots,m_{i}$, $x_{ij} \sim N(5,1)$, $\gamma_{0}=-0.5$, $\beta_{0i}=1+b_{0i}$, $\beta_{1i}=0.2+b_{1i}$, $\delta_{i}=0.2+d_{i}$ and $\Psi_{i}$ is given by (\ref{segfunction}) with $\lambda_{i}=5+\ell_{i}$, $L_{1}=\underset{\mathrm{1\leq j \leq m_{i}}}{\text{min}}(\mathbf{x}_{i})$ and $L_{2}=\underset{\mathrm{1\leq j \leq m_{i}}}{\text{max}}(\mathbf{x}_{i})$, where $\mathbf{x}_{i}^{\top}=(x_{i1},\ldots,x_{im_{i}})$. In addition, we have that $(b_{0i}, b_{1i}, d_{i},l_{i})^{\top}\sim  N(0, \mbox{diag}(0.1^{2}, 0.1^{2}, 0.1^{2}, 0.1^{2}))$. Finally, $m_{i} \equiv m \in (5, 20, 40, 160)$.

\textbf{Scenario 2:} The data were generated from the segmented ZIP mixed-effects model (\ref{zipmixed1}) with parameters simulated by the following linear predictors
\begin{eqnarray}
\text{log}(\mu_{ij}) &=& \beta_{0i}+\beta_{1i}x_{ij}+\delta_{i}(x_{ij}-\Psi_{i})_{+}^{2} \nonumber \\
\text{logit}(\pi_{ij}) &=& \gamma_{0}, \nonumber
\end{eqnarray}
where $i=1,\ldots,2000$, $j=1,\ldots,m_{i}$, $x_{ij} \sim N(5,1)$, $\gamma_{0}=0.5$, $\beta_{0i}=1+b_{0i}$, $\beta_{1i}=-0.2+b_{1i}$, $\delta_{i}=-0.2+d_{i}$ and $\Psi_{i}$ is given by (\ref{segfunction}) with $\lambda_{i}=2+\ell_{i}$. The quantities $L_{1}$, $L_{2}$, $m_{i}$, and the distribution of the vector $(b_{0i}, b_{1i}, d_{i}, l_{i})^{\top}$ are defined in the same way as in the scenario 1. 

Table \ref{zipsimutable} shows the empirical means, medians, standard deviations and absolute relative biases (ARB) for the point estimators of all the model parameters based on 500 replicates for scenarios 1 and 2 respectively. For each replicate we generate a grid of values varying from $L_{1}$ to $L_{2}$ with a step size of $0.25$ and the initial values for the changepoint obtained by grid search.

The behaviour of the estimators resulting from the proposed framework is satisfactory, each estimator appears to be approximately unbiased with variability decreasing as the sample size increases.

\begin{table}
\footnotesize\renewcommand{\arraystretch}{1}
\begin{tabular}{lccccc|cccc}
\toprule
& &\multicolumn{4}{c}{\textbf{Scenario 1}: Linear} &  \multicolumn{4}{c}{\textbf{Scenario 2}: Quadratic}\\
\cmidrule(lr){3-6} \cmidrule(lr){7-10} 

$m$ & Fixed effects & Mean & Median & sd & ARB ($\%$) & Mean & Median & sd & ARB ($\%$)\\
\midrule
$5$ & $\beta_{0}$ & $1.1165$ & $1.033$ & $0.413$ & $11.7\%$ & $0.9453$ & $0.971$ & $0.1022$ & $5.5\%$ \\
& $\beta_{1}$ & $0.1862$ & $0.197$ & $0.180$ & $6.9\%$ & $-0.2254$ & $-0.211$ & $0.0456$ & $12.7\%$ \\
& $\delta$ & $0.1808$ & $0.192$ & $0.180$ & $9.6\%$ & $-0.1867$ & $-0.193$ & $0.0410$ & $6.7\%$ \\
& $\lambda$ & $5.0836$ & $5.044$ & $0.580$ & $1.7\%$ & $1.8735$ & $1.960$ & $0.0791$ & $6.3\%$ \\
& $\gamma_{0}$ & $-0.4851$ & $-0.493$ & $0.207$ & $3.0\%$ & $0.5120$ & $0.506$ & $0.0207$ & $2.4\%$ \\
\midrule
$20$ & $\beta_{0}$ & $0.9867$ & $0.992$ & $0.361$ & $1.3\%$ & $0.9892$ & $0.994$ & $0.0415$ & $1.1\%$ \\
& $\beta_{1}$ & $0.2054$ & $0.204$ & $0.147$ & $2.7\%$ & $-0.1956$ & $-0.197$ & $0.0102$ & $2.2\%$ \\
& $\delta$ & $0.1968$ & $0.199$ & $0.154$ & $1.6\%$ & $-0.1927$ & $-0.196$ & $0.0046$ & $3.7\%$ \\
& $\lambda$ & $5.0152$ & $5.012$ & $0.358$ & $0.3\%$ & $1.9701$ & $1.986$ & $0.0309$ & $1.5\%$ \\
& $\gamma_{0}$ & $-0.5095$ & $-0.505$ & $0.103$ & $1.9\%$ & $0.4980$ & $0.499$ & $0.0103$ & $0.4\%$ \\
\midrule
$40$ & $\beta_{0}$ & $1.0121$ & $1.004$ & $0.348$ & $1.2\%$ & $1.0127$ & $1.004$ & $0.0289$ & $1.3\%$ \\
& $\beta_{1}$ & $0.1946$ & $0.197$ & $0.106$ & $2.7\%$ & $-0.1974$ & $-0.199$ & $0.0073$ & $1.3\%$ \\
& $\delta$ & $0.1981$ & $0.199$ & $0.111$ & $1.0\%$ & $-0.1957$ & $-0.198$ & $0.0032$ & $2.1\%$ \\
& $\lambda$ & $5.003$ & $5.001$ & $0.251$ & $0.1\%$ & $1.983$ & $1.993$ & $0.0201$ & $0.8\%$ \\
& $\gamma_{0}$ & $-0.4971$ & $-0.499$ & $0.105$ & $0.6\%$ & $0.4963$ & $0.499$ & $0.0073$ & $0.7\%$ \\
\midrule
$160$ & $\beta_{0}$ & $1.012$ & $1.005$ & $0.195$ & $1.2\%$ & $1.003$ & $1.001$ & $0.0146$ & $0.3\%$ \\
& $\beta_{1}$ & $0.201$ & $0.201$ & $0.055$ & $0.5\%$ & $-0.1993$ & $-0.200$ & $0.0041$ & $0.4\%$ \\
& $\delta$ & $0.2003$ & $0.200$ & $0.058$ & $0.1\%$ & $-0.1972$ & $-0.199$ & $0.0024$ & $1.4\%$ \\
& $\lambda$ & $4.999$ & $4.999$ & $0.122$ & $0.0\%$ & $1.9991$ & $1.999$ & $0.0099$ & $0.0\%$ \\
& $\gamma_{0}$ & $-0.5009$ & $-0.500$ & $0.036$ & $0.2\%$ & $0.4991$ & $0.500$ & $0.0036$ & $0.2\%$ \\
\bottomrule
\end{tabular}
\caption{Mean, median, standard deviation (values $\times$ 10), and absolute relative bias (ARB) of the sampling distributions for the parameter estimators in the segmented ZIP mixed-effects model with random changepoint for the two scenarios.} \label{zipsimutable}
\end{table}

\section{Applications} \label{sec5}
\subsection{Known changepoint case}
The parameter estimates from the segmented ZIP mixed-effects model (\ref{covid1}) with an unstructured covariance matrix are reported in Table \ref{segzip_fixed_fixeff}. Regarding the log-linear model, we can observe that all fixed effects were statistically significant, with emphasis on the segmentation function, providing statistical evidence that the COVID-19 pandemic increased the overall CR-BSI rate. Specifically, based on the estimated model, before the pandemic, the expected CR-BSI rate decreased by $100\times (1-\text{exp}(-0.007))\approx 0.70\%$ per month, and immediately after the onset of the pandemic ($t_{ij}>39$), the expected CR-BSI rate increased by $100\times (\text{exp}(0.039-0.007))-1\approx 3.25\%$ per month, holding other variables constant, consistent with the exploratory data analysis. If the hospital is private, the CR-BSI rate is expected to be approximately $100\times (1-\text{exp}(-0.841))\approx 56.87\%$ lower compared to non-private hospitals, as observed in the EDA.

Regarding the logit model, the type of hospital has a significant effect, suggesting that the incidence of zeros (absence of CR-BSI) is higher in private hospitals, as seen in the EDA section.

\begin{table}
\centering
\small\renewcommand{\arraystretch}{0.8}
\begin{tabular}{lcccc}
\toprule
\bf{Log-linear model}& Parameter &  Estimate & Std. Error & $p$-value\\
\midrule
Fixed Effects:& &   &  & \\
\hspace{0.2cm}Intercept& $\beta_{0}$ & $-5.345$  & $0.117$ & $<0.001$\\
\hspace{0.2cm}Time (Month)& $\beta_{1}$ & $-0.007$  & $0.003$ & $<0.001$ \\
\hspace{0.2cm}Type (Private)& $\beta_{2}$ & $-0.841$  & $0.152$ & $0.017$ \\
\hspace{0.2cm}COVID-19& $\delta$ & $0.039$ & $0.017$ & $0.020$\\
Variance Components ($\mathbf{G}$):& &   &  & \\
\hspace{0.2cm}$\mathbb{V}(b_{0i})$& $\sigma_{b_{0}}^{2}$ & $0.404$&  & \\
\hspace{0.2cm}$\mathbb{V}(b_{1i})$& $\sigma_{b_{1}}^{2}$ & $0.0003$ &  & \\
\hspace{0.2cm}$\mathbb{V}(d_{i})$& $\sigma_{d}^{2}$ & $0.007$ & & \\
\hspace{0.2cm}$\text{cov}(b_{0i},b_{1i})$& $\sigma_{b_{0}b_{1}}$ & $-0.489$  &  & \\
\hspace{0.2cm}$\text{cov}(b_{0i},d_{i})$& $\sigma_{b_{0}d}$ &  $0.103$ &  & \\
\hspace{0.2cm}$\text{cov}(b_{1i},d_{i})$& $\sigma_{b_{1}d}$ &  $-0.641$ &  & \\
Systemic effect variance:& &   &  & \\
\hspace{0.2cm}$\mathbb{V}(\zeta_{t_{ij}})$& $\sigma_{\zeta}^{2}$ & $0.005$  &  & \\
\toprule
\bf{Zero-inflation model}& &   &  & \\
\midrule
\hspace{0.2cm}Intercept& $\gamma_{0}$ & $-5.662$ & $1.139$ & $<0.001$ \\
\hspace{0.2cm}Type (Private)& $\gamma_{1}$ & $2.738$  & $1.280$ & $0.032$\\
\midrule
Log-likelihood&  & $-4000.1$  &  & \\
AIC& & $8026.1$&  & \\
BIC& & $8103.8$ &  & \\
Deviance&  &  $8000.1$ &  & \\
\bottomrule
\end{tabular}
\caption{Parameter estimates from segmented ZIP mixed-effects model with a known changepoint and unstructured covariance matrix.} \label{segzip_fixed_fixeff}
\end{table}

\begin{figure}
     \centering
     \begin{subfigure}[b]{0.45\textwidth}
         \centering
         \includegraphics[width=\textwidth]{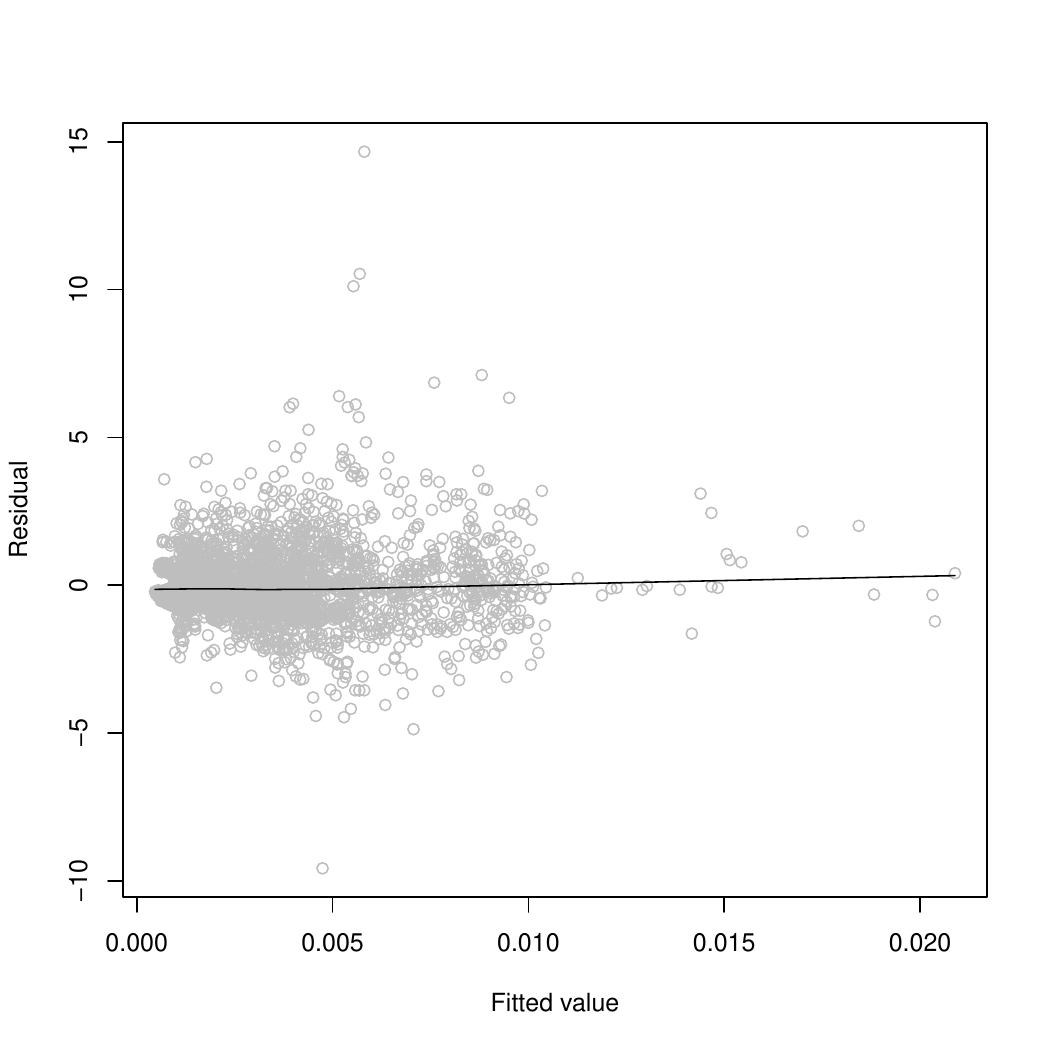}
         \caption{Residuals vs fitted values}
         \label{diagnostic_fixed1}
     \end{subfigure}
     \hfill
     \begin{subfigure}[b]{0.45\textwidth}
         \centering
         \includegraphics[width=\textwidth]{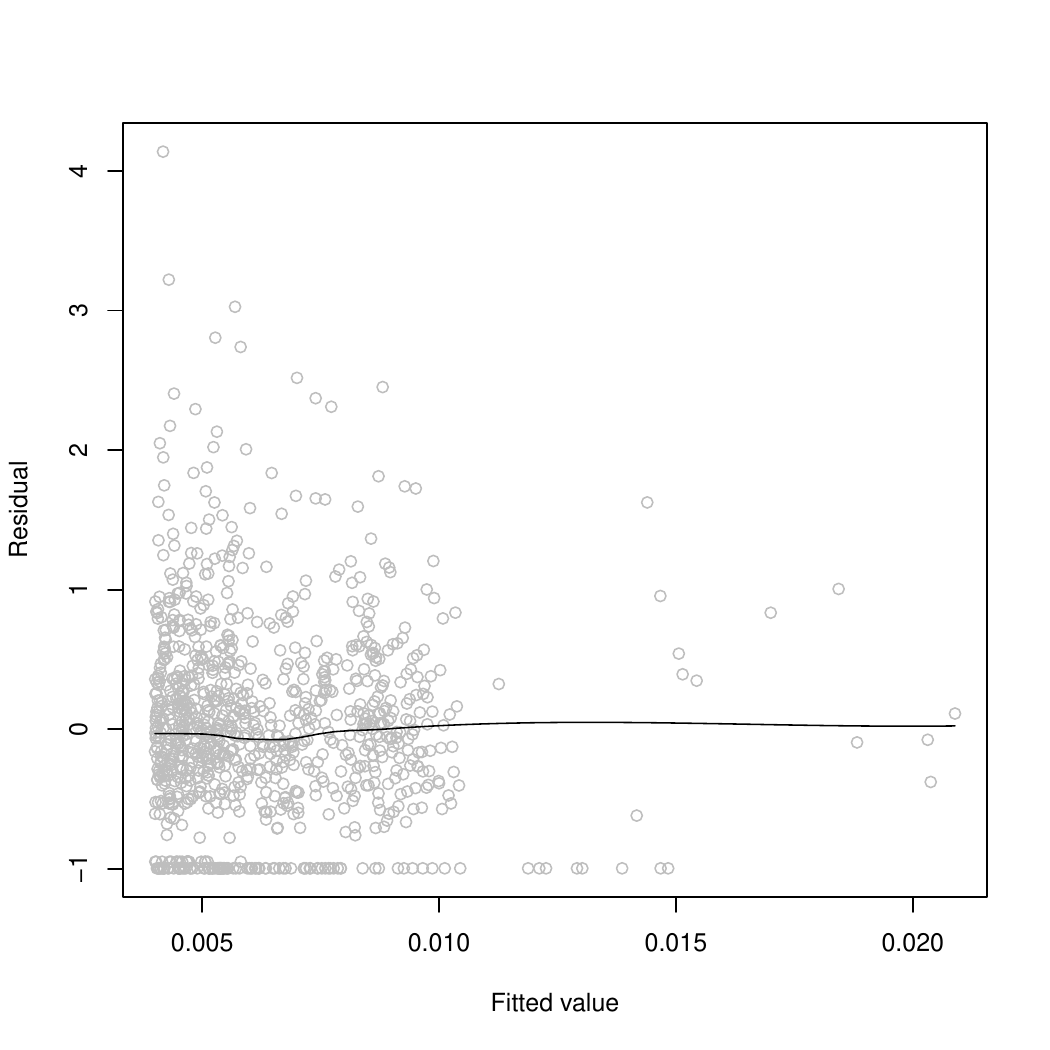}
         \caption{Working residuals vs fitted values}
         \label{diagnostic_fixed2}
     \end{subfigure}
     \vfill
     \begin{subfigure}[b]{0.48\textwidth}
         \centering
         \includegraphics[width=\textwidth]{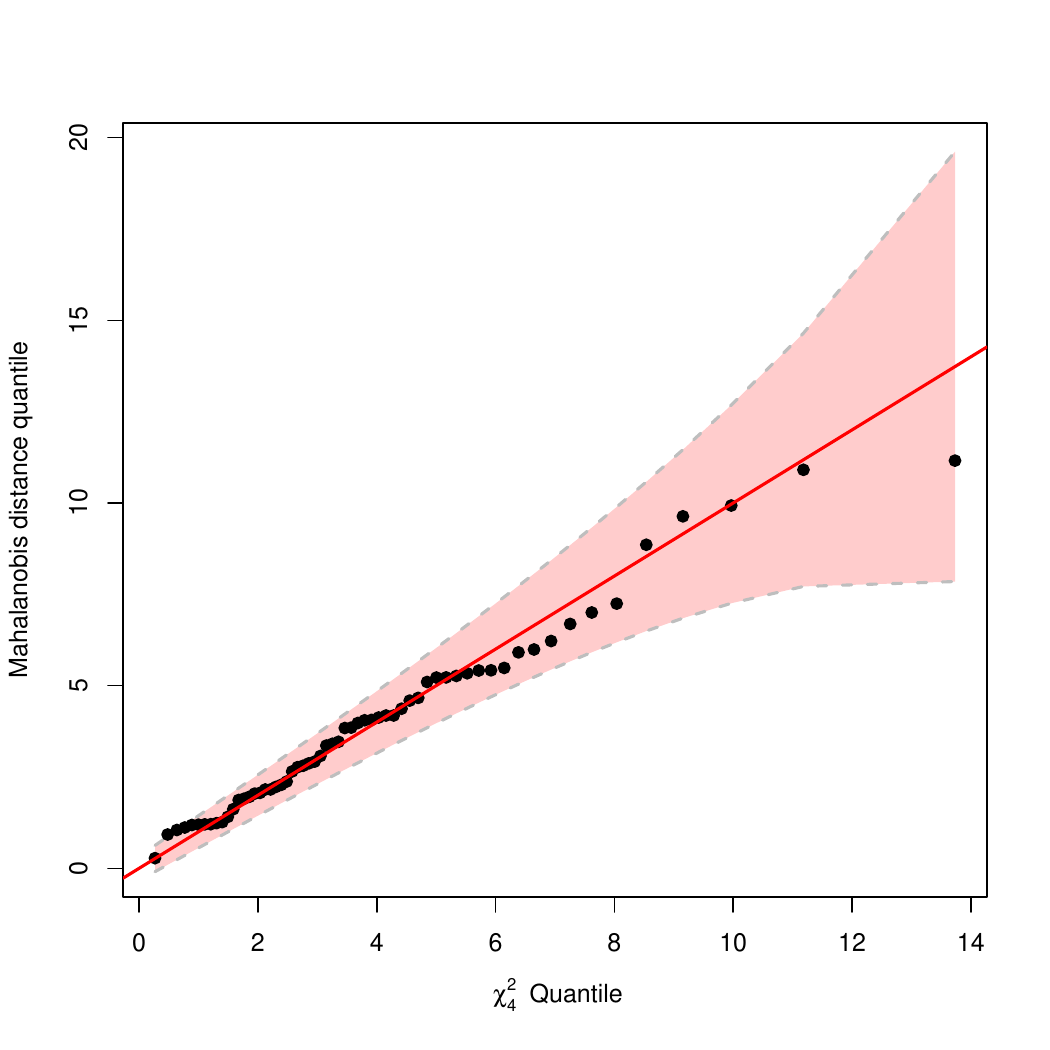}
         \caption{Mahalanobis distance QQ-plot for random effects}
         \label{diagnostic_fixed3}
     \end{subfigure}
     
        \caption{Diagnostic plot for the known changepoint case.}
        \label{diagnostic_fixed}
\end{figure}

Some diagnostic plots for the known changepoint case are depicted in Figure \ref{diagnostic_fixed} and do not exhibit abrupt deviations with respect to the linearity assumption of the linear predictor (Figure \ref{diagnostic_fixed1}) shows no evidence against the assumed  canonical link function. In particular, the Mahalanobis distance QQ-plot (Figure \ref{diagnostic_fixed3}) shows no evidence against the Gaussian assumption adopted for either random effects ($\mathbf{b}_{i}, \zeta_{t_{ij}})^{\top}$.

\begin{figure}
\centering
\begin{subfigure}{0.75\textwidth}
\includegraphics[width=\textwidth]{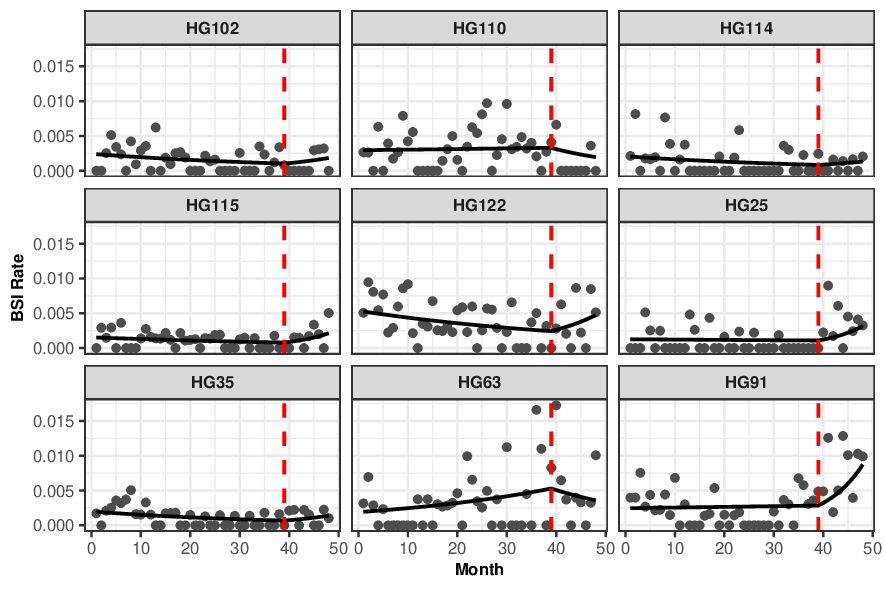}
\caption{Private hospitals}
\label{fig:ex3a}
\end{subfigure} 
\hfill
\begin{subfigure}{0.75\textwidth}
\includegraphics[width=\textwidth]{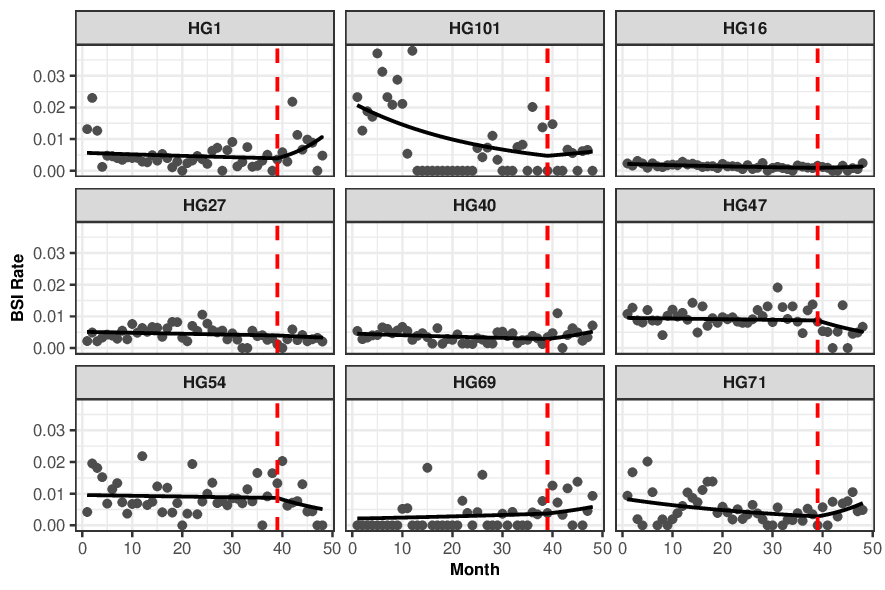}
\caption{Non-private hospitals}
\label{fig:ex3b}
\end{subfigure}
\caption{Observed data points for 9 selected hospitals along with fitted segmented ZIP mixed-effects model with known changepoint. The COVID-19 pandemic is represented by the dashed line ($t_{ij}=39$).}
\label{countdist1}
\end{figure}

The Figure \ref{countdist1} shows, for 9 selected hospitals, the observed data along with the fitted values from the estimated model for private and non-private hospitals, respectively. We can observe a good fit of the model to the data.

Finally, it is important to mention that the model \ref{covid1} with smooth transition $(t_{ij}-39)_{+}^{2}$ was also tested. However, the model with linear segmentation function showed better results: MSE of $1.58$, AIC of $8026.10$, and BIC of $8103.80$ compared to MSE of $1.59$, AIC of $8042.00$, and BIC of $8119.60$, respectively.

\subsection{Random changepoint case}
Let $Y_{ij}$ defined as in Section \ref{sec3_1}. Now we generalize the log-linear model specification of the equation (\ref{covid1}) by including a random changepoint, that is
\begin{align}
\label{covid2}
\begin{split}
&\text{log}(\mu_{ij})=\text{log}(O_{ij})+\beta_{0i}+\beta_{1i}t_{ij}+\delta_{i}(t_{ij}-\Psi_{i}(\lambda_{i}))_{+}+\beta_{2}N_{i}+\zeta_{t_{ij}} \\
&\text{logit}(\pi_{ij})= \gamma_{0}+\gamma_{1}N_{i}, 
\end{split} 
\end{align} 
with
\begin{eqnarray}
\Psi_{i}(\lambda_{i})=\frac{L_{1}+L_{2}\text{exp}(\lambda_{i})}{1+\text{exp}(\lambda_{i})}, \nonumber
\end{eqnarray}
where $O_{ij}$ is the offset correponding to the CVC, $t_{ij}$ represents the time (month), $N_{i}$ the type of $i$-th hospital and $\zeta_{t_{ij}}$ captures the systemic effect of the month ($t_{ij}$) in all hospitals. To complete the model specification, we assume that $\beta_{0i}=\beta_{0}+b_{0i}$, $\beta_{1i}=\beta_{1}+b_{1i}$, $\delta_{i}=\delta+d_{i}$ and $\lambda_{i}=\lambda+l_{i}$. We fix $L_{1}=1$ and $L_{2}=48$ and assume that the random effects $\mathbf{b}_{i}=(b_{0i}, b_{1i}, d_{i}, l_{i})^{\top}$ are independent of each other for $i=1,\ldots,61$ and of $\zeta_{t_{ij}}$ with
\begin{eqnarray}
\mathbf{b}_{i} &\sim & \mathbf{N}_{4}(\mathbf{0}, \mathbf{G}) \nonumber \\
\zeta_{t_{ij}} &\overset{\mathrm{iid}}{\sim} & N(0,\sigma_{\zeta}^{2}). \nonumber
\end{eqnarray} 
Now, the unstructured variance-covariance matrix may be expressed as
\begin{eqnarray}
\mathbf{G}=\begin{pmatrix}
\sigma_{b_{0}}^{2} & \sigma_{b_{0}b_{1}} & \sigma_{b_{0}d} & \sigma_{b_{0}l} \\
& \sigma_{b_{1}}^{2} & \sigma_{b_{1}d} & \sigma_{b_{1}l} \\
& & \sigma_{d}^{2}& \sigma_{d_{l}} \\
& & & \sigma_{l}^{2} 
\end{pmatrix} \label{covariance2}
\end{eqnarray}
For the selection of the initial changepoint, we used the grid search. The parameter estimates from the segmented ZIP mixed-effects model (\ref{covid2}) with unstructured covariance matrix are reported in Table \ref{segzip_random_fixeff}. Regarding the log-linear model, we can observe that all fixed effects are statistically significant, with emphasis on the segmentation function, which provides statistical evidence that the CR-BSI rate generally increases after the regime change. More specifically, based on the estimated model, prior to the changepoint, the expected CR-BSI rate decreases by approximately $1.30\%$ per month ($100\times (1-\text{exp}(-0.013))$), and immediately after, it is observed that the expected CR-BSI rate increases by approximately $0.90\%$ per month ($100\times (\text{exp}(0.022-0.013))-1$), holding other variables constant. If the hospital is private, the CR-BSI rate is expected to be approximately $61.02\%$ lower compared to non-private hospitals, as observed in the EDA. As for the logit model, the inference is the same as in the known changepoint case. 

The estimate for the population change point is $\hat{\Psi}=(1+48\text{exp}(0.274))/(1+\text{exp}(0.274))\approx 27.70$, which corresponds to a period between March 2019 and April 2019. Considering Figure 3 of the Supplementary Material, we can see that the point of transition occurs near the month $t_{ij}=29$ (May 2019). Therefore, the estimated value $\hat{\Psi}$ is very close to the observed value in the EDA.

\begin{table}
\centering
\small\renewcommand{\arraystretch}{0.8}
\begin{tabular}{lcccc}
\toprule
\bf{Log-linear model}& Parameter &  Estimate & Std. Error & $p$-value\\
\midrule
Fixed Effects:& &   &  & \\
\hspace{0.2cm}Intercept& $\beta_{0}$ & $-5.275$  & $0.119$ & $<0.001$\\
\hspace{0.2cm}Time (Month)& $\beta_{1}$ & $-0.013$  & $0.004$ & $<0.001$ \\
\hspace{0.2cm}Type (Private)& $\beta_{2}$ & $-0.942$  & $0.152$ & $0.024$ \\
\hspace{0.2cm}Diff slope& $\delta$ & $0.022$ & $0.017$ & $0.015$\\
\hspace{0.2cm}Changepoint& $\lambda$ & $0.274$ & $0.012$ & $0.030$\\
Variance Components ($\mathbf{G}$):& &   &  & \\
\hspace{0.2cm}$\mathbb{V}(b_{0i})$& $\sigma_{b_{0}}^{2}$ & $0.455$&  & \\
\hspace{0.2cm}$\mathbb{V}(b_{1i})$& $\sigma_{b_{1}}^{2}$ & $0.0007$ &  & \\
\hspace{0.2cm}$\mathbb{V}(d_{i})$& $\sigma_{d}^{2}$ & $0.004$ & & \\
\hspace{0.2cm}$\mathbb{V}(l_{i})$& $\sigma_{l}^{2}$ & $0.273$ & & \\
\hspace{0.2cm}$\text{cov}(b_{0i},b_{1i})$& $\sigma_{b_{0}b_{1}}$ & $-0.548$  &  & \\
\hspace{0.2cm}$\text{cov}(b_{0i},d_{i})$& $\sigma_{b_{0}d}$ &  $0.394$ &  & \\
\hspace{0.2cm}$\text{cov}(b_{0i},l_{i})$& $\sigma_{b_{0}l}$ &  $0.088$ &  & \\
\hspace{0.2cm}$\text{cov}(b_{1i},d_{i})$& $\sigma_{b_{1}d}$ &  $-0.865$ &  & \\
\hspace{0.2cm}$\text{cov}(b_{1i},l_{i})$& $\sigma_{b_{1}l}$ &  $0.250$ &  & \\
\hspace{0.2cm}$\text{cov}(d_{i},l_{i})$& $\sigma_{dl}$ &  $-0.397$ &  & \\
Systemic effect variance:& &   &  & \\
\hspace{0.2cm}$\mathbb{V}(\zeta_{t_{ij}})$& $\sigma_{\zeta}^{2}$ & $0.004$  &  & \\
\toprule
\bf{Zero-inflation model}& &   &  & \\
\midrule
\hspace{0.2cm}Intercept& $\gamma_{0}$ & $-5.564$ & $1.067$ & $<0.001$ \\
\hspace{0.2cm}Type (Private)& $\gamma_{1}$ & $2.348$  & $1.800$ & $0.072$\\
\midrule
Log-likelihood&  & $-3995.4$  &  & \\
AIC& & $8016.4$&  & \\
BIC& & $8093.2$ &  & \\
Deviance&  &  $7994.2$ &  & \\
\bottomrule
\end{tabular}
\caption{Parameter estimates from the segmented ZIP mixed-effects model with random changepoint with unstructured covariance matrix.} \label{segzip_random_fixeff}
\end{table} 

Diagnostic plots for the random changepoint case show a similar picture compared to the known case and are displayed in Figure 8, Section 2 of Supplementary Material.

The Figure \ref{countdist1_random} illustrates, for 9 selected hospitals, the observed data along with the fitted values from the estimated model for private and non-private hospitals, respectively. We can observe a good fit of the model to the observed data.

\begin{figure}
\centering
\begin{subfigure}{0.75\textwidth}
\includegraphics[width=\textwidth]{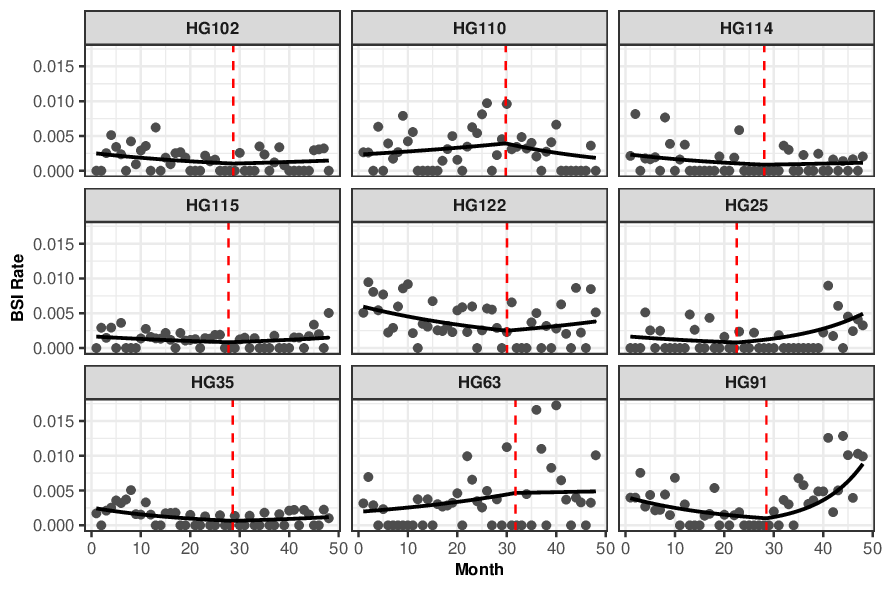}
\caption{Private hospitals}
\label{fig:exa}
\end{subfigure} 
\vfill
\begin{subfigure}{0.75\textwidth}
\includegraphics[width=\textwidth]{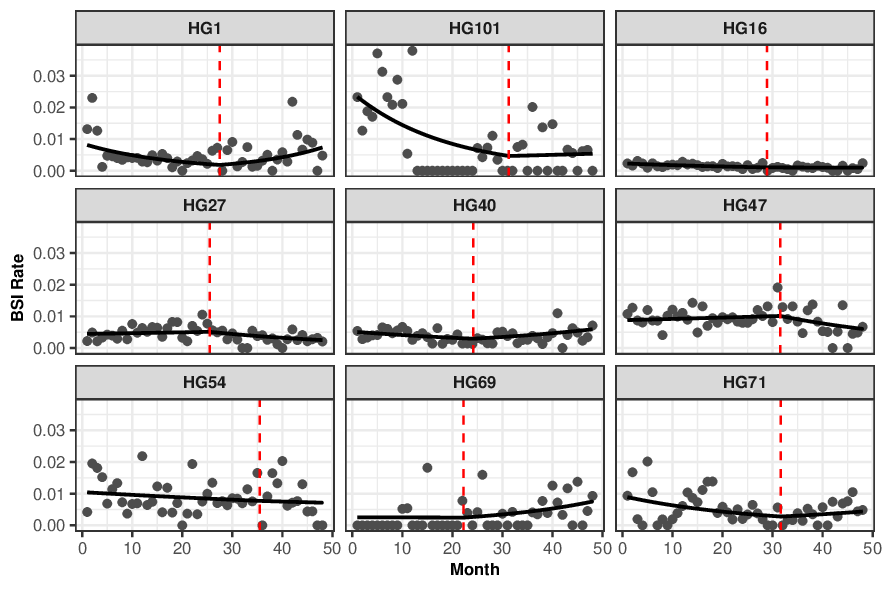}
\caption{Non-private hospitals}
\label{fig:exb}
\end{subfigure}
\caption{Observed data points for 9 selected hospitals along with fitted segmented ZIP mixed-effects model with random changepoint. The dashed line represents the estimated changepoint.}
\label{countdist1_random}
\end{figure}

Comparing models (\ref{covid2}) and (\ref{covid1}), the mean squared errors were $1.54$ and $1.58$ respectively; the information criteria as well as the deviance show that the model with a random changepoint gives better results when compared to the model with a known changepoint. Thus, the model allows for greater flexibility to adapt to the regime changes of each hospital, which results in a gain in performance.

\section{Discussion} \label{sec6}
According to P\' erez-Granda et al. (2022), information regarding the impact of the COVID-19 pandemic on the incidence of catheter-related bloodstream infections (CR-BSIs) is currently limited. Therefore, in this article, we propose a methodology to investigate the effect of the COVID-19 pandemic on CR-BSI based on segmented zero-inflated Poisson (ZIP) mixed-effects models. First, the segmented ZIP mixed-effects model with a known changepoint in March 2020 was fitted to the CR-BSI data from private and non-private hospitals located in the city of S\~ao Paulo. The results show that the COVID-19 pandemic contributed to an increase in the CR-BSI rate. In addition, it was observed that private hospitals had a lower average CR-BSI rate compared to non-private hospitals.

Based on simulated data, we evaluated the efficiency of our algorithm in two different scenarios. The estimators derived from the proposed framework exhibit satisfactory behaviour, appearing to be approximately unbiased with reduced variability for larger sample sizes. After the simulation exercises, the model was fitted to the motivational data. When considering the random changepoint, an improvement in CR-BSI prediction was observed compared to the model with a known changepoint. From a practical perspective, the proposed model may assist researchers in identifying other points at which changes in CR-BSI regimes occurred, allowing for an assessment of the reasons for such changes beyond the COVID-19 pandemic alone.

The algorithm for estimating the segmented ZIP mixed model with a random changepoint demonstrated simplicity and efficiency, but it has the following main limitations: 1) The algorithm is highly sensitive to the choice of the initial changepoint $\Psi_{0}$. 2) In this article, we used grid-search to select $\Psi_{0}$, which made the estimation process computationally expensive.

To improve the robustness and efficiency of the algorithm, the following ideas are under development: 1) Implementation of the bootstrap restarting (Wood, 2001) to select the initial value of the changepoint $\Psi_{0}$. 2) Inclusion of covariates in the changepoint equation (\ref{segfunction}) to assess their impact on the changepoint location. 3) Development of model diagnostic procedures based on various residuals for mixed-effects generalized models, inspired by the work of Singer et al. (2017).


\begin{thebibliography}{99}
\bibitem[Brooks et al. (2017)]{Brooks} Brooks M.E., Kristensen K., van Benthem K.J., Magnusson A., Berg CW, Nielsen A., Skaug H.J., Maechler M. and Bolker B.M. (2017). glmmTMB balances speed and flexibility among packages for zero-inflated generalized linear mixed modeling. {\em The R Journal, 9(2)}, 378--400.

\bibitem[Das et al. (2016)]{Das} Das R., Banerjee M., Nan B. and Zheng H. (2016). Fast estimation of regression parameters in a broken-stick model for longitudinal data. {\em Journal of the American Statistical Association, 111}, 1132--1143. 

\bibitem[Feder (1975)]{Feder} Feder P. (1975). On asymptotic distribution theory in segmented regression problems-identified cases. {\em The Annals of Statistics, 3}, 49--83.

\bibitem[Fitzmaurice et al. (2001)]{Fitzmaurice} Fitzmaurice G.M., Laird N.M, and Ware J.H (2004). Applied Longitudinal Analysis. New York: Wiley. 

\bibitem[Gahlot et al. (2014)]{Gahlot} Gahlot R., Nigam C., Kumar V., Yadav G. and Aupurba S. (2014).Catheter-related bloodstream infections. {\em Int J Crit Illn Inj Sci, 4(2)}, 162--167.

\bibitem[Ghosh and Vaida (2007)]{Ghosh} Ghosh P., and Vaida F. (2007). Random changepoint modelling of HIV immunological responses. {\em Statistics in Medicine, 26}, 2074--87.

\bibitem[Hall et al. (2000)]{Hall} Hall C.B., Lipton R.B., Sliwinski M. and Stewart W.F. (2000). A change point model for estimating the onset of cognitive decline in preclinical Alzheimer's disease. {\em Statistics in Medicine, 19}, 1555--66.

\bibitem[Hall (2000)]{Hall1} Hall D.B. (2000). Zero-inflated poisson and binomial regression with random effects: a case study. {\em Biometrics, 56(4)}, 1030--1039.



\bibitem[Jacqmin-Gadda et al. (2006)]{Jacqmin} Jacqmin-Gadda H., Commenges D., and Dartigues J.F. (2006). Random changepoint model for joint modeling of cognitive decline and dementia. {\em Biometrics, 62}, 254--60.

\bibitem[Jiang (2017)]{Jiang} Jiang J. (2017). Asymptotic Analysis of Mixed Effects Models: Theory, Applications, and Open Problems. Chapman \& Hall - CRC Press.

\bibitem[La Cruz et al. (2006)]{La Cruz} La Cruz W., Martinez J.M. and Raydan M. (2006). Spectral residual method without gradient information for solving large-scale non-linear systems of equations. {\em Mathematics of Computation, 75,} 1429--1448.

\bibitem[Lerman (1980)]{Lerman} Lerman P.M. (1980). Segmented regression models by grid search. {\em Journal of the Royal Statistical Society. Series C (Applied Statistics)}, 29, 77--84.

\bibitem[Lai and Albert (2014)]{Lai} Lai Y. and Albert P.S. (2014). Identifying multiple change points in a linear mixed effects model. {\em Statistics in Medicine, 33}, 1015--1028.

\bibitem[Lambert (1992)]{Lambert} Lambert D. (1992). Zero-inflated Poisson regression, with an application to defects in manufacturing. {\em Technometrics, 34}, 1--13.

\bibitem[Liang and Zeger (2016)]{Liang} Liang K.Y. and Zeger S.L. (1986). Longitudinal data analysis using generalized linear models. {\em Biometrika, 73}, 13--22.

\bibitem[Liu et al. (2020)]{Liu} Liu J., Ma Y. and Johnstone Y.Q. (2020). A Goodness-of-fit Test for Zero-Inflated Poisson Mixed Effects Models in Tree Abundance Studies. {\em Computational Statistics and Data Analysis, 144}.

\bibitem[Muggeo et al. (2014)]{Muggeo} Muggeo V.M.R., Atkins D.C. and Gallop R.J. (2014). Segmented mixed models with random changepoints: a maximum likelihood approach with application to treatment for depression study. {\em Statistical Modelling, 14}, 293--313.

\bibitem[P\' erez-Granda et al. (2022)]{Perez-Granda} P\' erez-Granda M.D., Carrillo C.S., Rabad\` an P.M., Valerio M., Olmedo M., Mu\~noz P. and Bouza E. (2022).Increase in the frequency of catheter-related bloodstream infections during the COVID-19 pandemic: a plea for control. {\em J Hosp Infect, 119}, 149--154.

\bibitem[Scott et al. (2004)]{Scott} Scott M.A., Norman R.G. and Berger K. (2004). Modelling growth and decline in lung function in Duchenne's muscolar dystrophy with an augmented linear mixed effects model. {\em Applied Statistics, 53}, 507--21.

\bibitem[Segalas et al. (2019)]{Segalas} Segalas C, Amieva H and Jacqmin-Gadda H. (2019). A hypothesis testing procedure for random changepoint mixed models. {\em Statistics in Medicine, 38}, 3791--3803.

\bibitem[Singer et al. (2017)]{Singer1} Singer J.M., Rocha F.M.M. and Nobre J.S. (2017). . Graphical tools for detecting departures from linear mixed models assumptions and some remedial measures. {\em International Statistical Review, 85}, 290--324.

\bibitem[Singer et al. (2020)]{Singer} Singer J.M., Rocha F.M.M., Pedroso-de-Lima A.C., Silva G.L., Coatti G.C. and Zatz M. (2020). Random changepoint segmented regression with smooth transition. {\em Statistical Methods in Medical Research, 0(0)}, 1--12.

\bibitem[Su et al. (2021)]{Su} Su W., Gecili E., Wang X. and Szczesniak R.D. (2021). An empirical comparison of segmented and stochastic linear mixed effects models to estimate rapid disease progression in longitudinal biomarker studies. {\em Stat Biopharm Res 13(3)}, 270--279.

\bibitem[Tapsoba et al. (2020)]{Tapsoba} Tapsoba J.D., Wang C-Y., Zangeneh S. and Chen Y.Q. (2020). Methods for Generalized Change-point Models: with  Applications to HIV Surveillance and Diabetes Data. {\em Statistics in Medicine, 39(8)}, 1167--1182.

\bibitem[Wood (2001)]{Wood} Wood S.N. (2001). Minimizing model fitting objectives that contain spurious local  minima by bootstrap restarting. {\em Biometrics, 57}, 240-44.

\bibitem[Zhou et al. (2020)]{Zhou} Zhou Z., Gao Y.,Yao W. and Yu N. (2020). A Robust Segmented Mixed Effect Regression Model for Baseline Electricity Consumption Forecasting. {\em Journal of Modern Power Systems and Clean Energy, 10(1)}, 71--80.
\end{thebibliography}
\end{document}


\def\spacingset#1{\renewcommand{\baselinestretch}%
{#1}\small\normalsize} \spacingset{1}


\bigskip
\begin{center}
{\large\bf SUPPLEMENTARY MATERIAL}
\end{center}

\section{Explanatory Data Analysis - EDA}
The purpose of this section is to identify possible patterns that may help to specify the model to be developed in the paper, as well as to raise some hypotheses of interest.

A preliminary analysis of the dataset, identified $121$ observations in which the CVC was zero. Since, as there was no use of the catheter and consequently no CR-BSI, so this information was removed from the dataset prior to exploratory analysis.

Table \ref{descritiva} shows some descriptive statistics for the $N = 76$ hospitals analyzed in this paper.

\begin{table}[H]
\centering
\begin{tabular}{l|c|c} 
\toprule
&\multicolumn{1}{c}{Private} &  \multicolumn{1}{c}{Non-private}\\
\midrule
No. of hospitals  & $34$  & $42$  \\
($N=76$) &  &   \\
 &  &  \\
No. of observations & $1,600$ & $1,866$ \\
($Total=3,466$) &  &   \\
 &  &    \\
BSI (Count) &  &  \\
Sum & $1,240$ & $3,856$\\
 &  &    \\
CVC (Count) &  &   \\
Sum & $658,527$  & $832,779$ \\
 &  &   \\
BSI rate &  &  \\
 $100\times(BSI/CVC)$ (range)& $0.188\%$ ($0-9.09\%$) & $0.463\%$ ($0-11.1\%$)\\
\bottomrule
\end{tabular} 
\caption{Some descriptive analysis of CR-BSI motivation data} \label{descritiva}
\end{table}

\begin{figure}
\centering
\includegraphics[width=16cm,height=8cm]{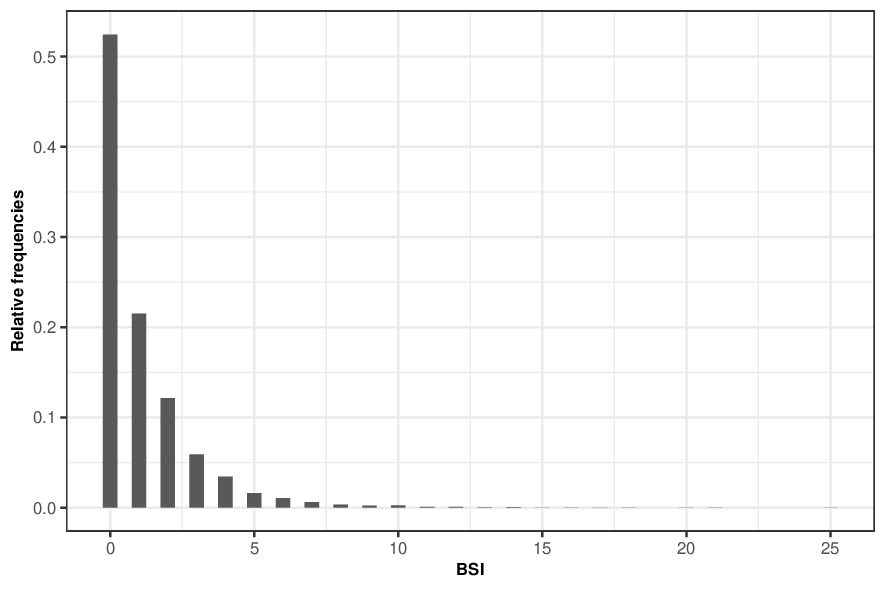}
\caption{CR-BSI count distribution.}
\label{countdist11}
\end{figure}

\begin{figure}
\centering
\includegraphics[width=16cm,height=8cm]{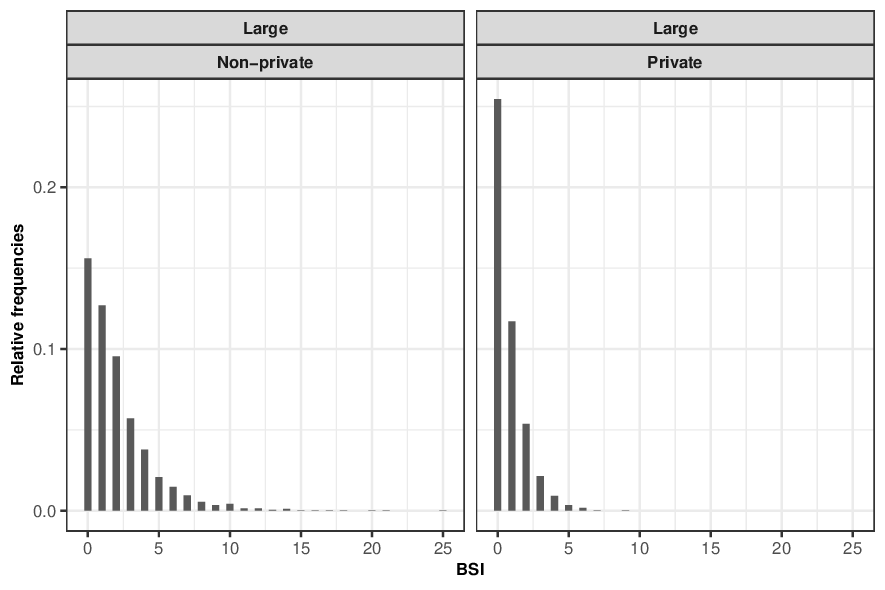}
\caption{CR-BSI count distribution grouped by type.}
\label{countdist22}
\end{figure}

The Figure \ref{countdist11} shows the frequency distribution of the number of CR-BSIs throughout the observation period, highlighting the high frequency of zeros in the data. Therefore, it seems reasonable to use the ZIP model to describe the evolution of the CR-BSI rate. Figure \ref{countdist22} shows the frequency distribution of the number of CR-BSIs separated by type. We can observe that the frequency of zeros is higher in private hospitals. Therefore, it is expected that the chance of BSI occurrence in private hospitals is lower when compared to non-private hospitals.

\begin{figure}[h]
\centering
\includegraphics[width=16cm,height=8cm]{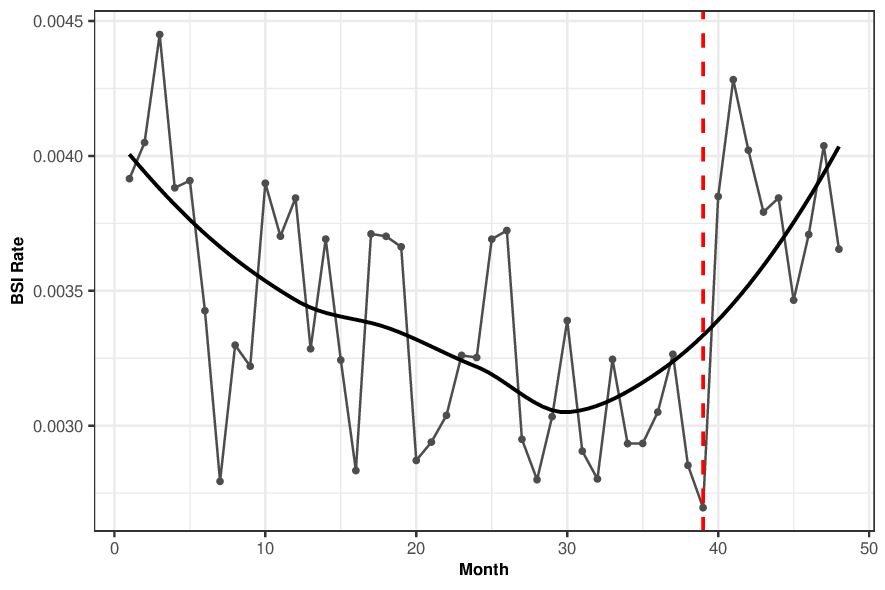}
\caption{Aggregated CR-BSI rate time series along with the LOESS curve. The COVID-19 pandemic is represented by the dashed line.}
\label{bsitime}
\end{figure}

The Figure \ref{bsitime} shows the evolution of the CR-BSI rate over time, along with the LOESS curve, aggregating all hospitals per month, i.e., $R_{j}^{*}=\sum_{i=1}^{N}BSI_{ij}/\sum_{i=1}^{N}CVC_{ij}$, $j=1,\ldots, 48$, where $R_{j}^{*} \in (0,1)$. It is worth noting that the average rate (loess curve) consistently decreases until about May 2019 ($t_{ij}=29$), and then exhibits an increasing trend. It is interesting to observe that shortly after the onset of the pandemic (dashed line), there was a jump in the BSI rate, indicating a possible regime change in the time series. This suggests that the COVID-19 pandemic has increased the incidence of BSIs in hospitals in São Paulo, in general.

\begin{figure}
\centering
\includegraphics[width=16cm,height=8cm]{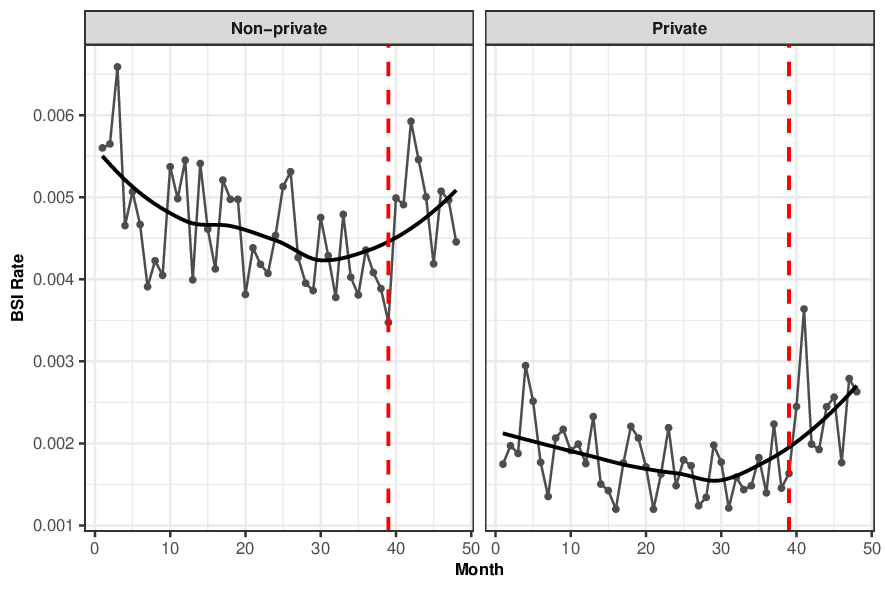}
\caption{Aggregated CR-BSI rate time series along with the LOESS curve grouped by type of hospital. The COVID-19 pandemic is represented by the dashed line.}
\label{bsitime_nat}
\end{figure}

The Figures \ref{bsitime_nat} shows the evolution of the aggregated CR-BSI rate over time, along with the LOESS curve, grouped by hospital type. The same overall trend is observed, and it can be seen that the CR-BSI rate is higher in non-private hospitals. 

\begin{figure}
    \includegraphics[width=8cm,height=8cm]{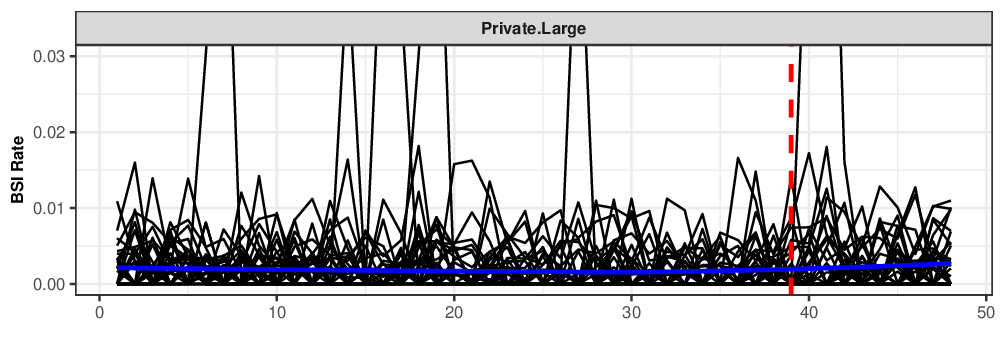}
    \includegraphics[width=8cm,height=8cm]{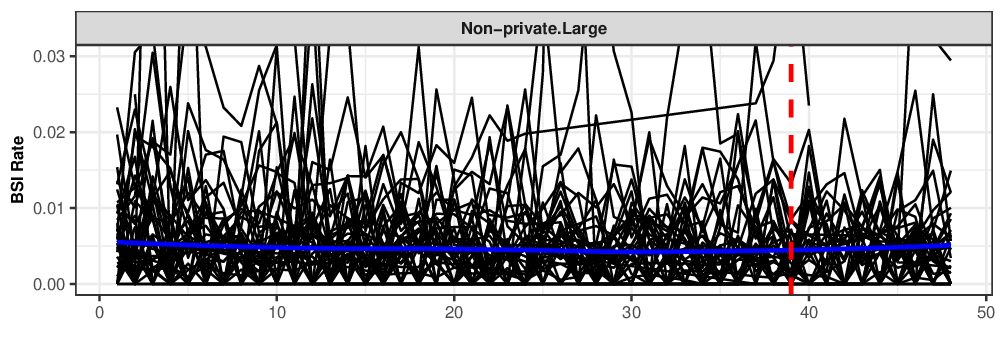}
    \caption{Profile plot for BSI rate along with the LOESS curve.}
     \label{bsiprofile}
\end{figure}

The profile plots for the CR-BSI rates with the LOESS curve are shown in Figure \ref{bsiprofile}. For both types of the hospitals, the following patterns can be observed: 1) high volatility in CR-BSI rates, with multiple peaks; 2) the trend in CR-BSI rates decreases consistently until approximately June 19 ($t_{ij}=30$) and shows an increasing trend, possibly further increasing after the onset of the COVID-19 pandemic, indicating a possible changepoint. Thus, descriptively there is at least one changepoint.

\begin{figure}
    \includegraphics[width=17cm,height=17cm]{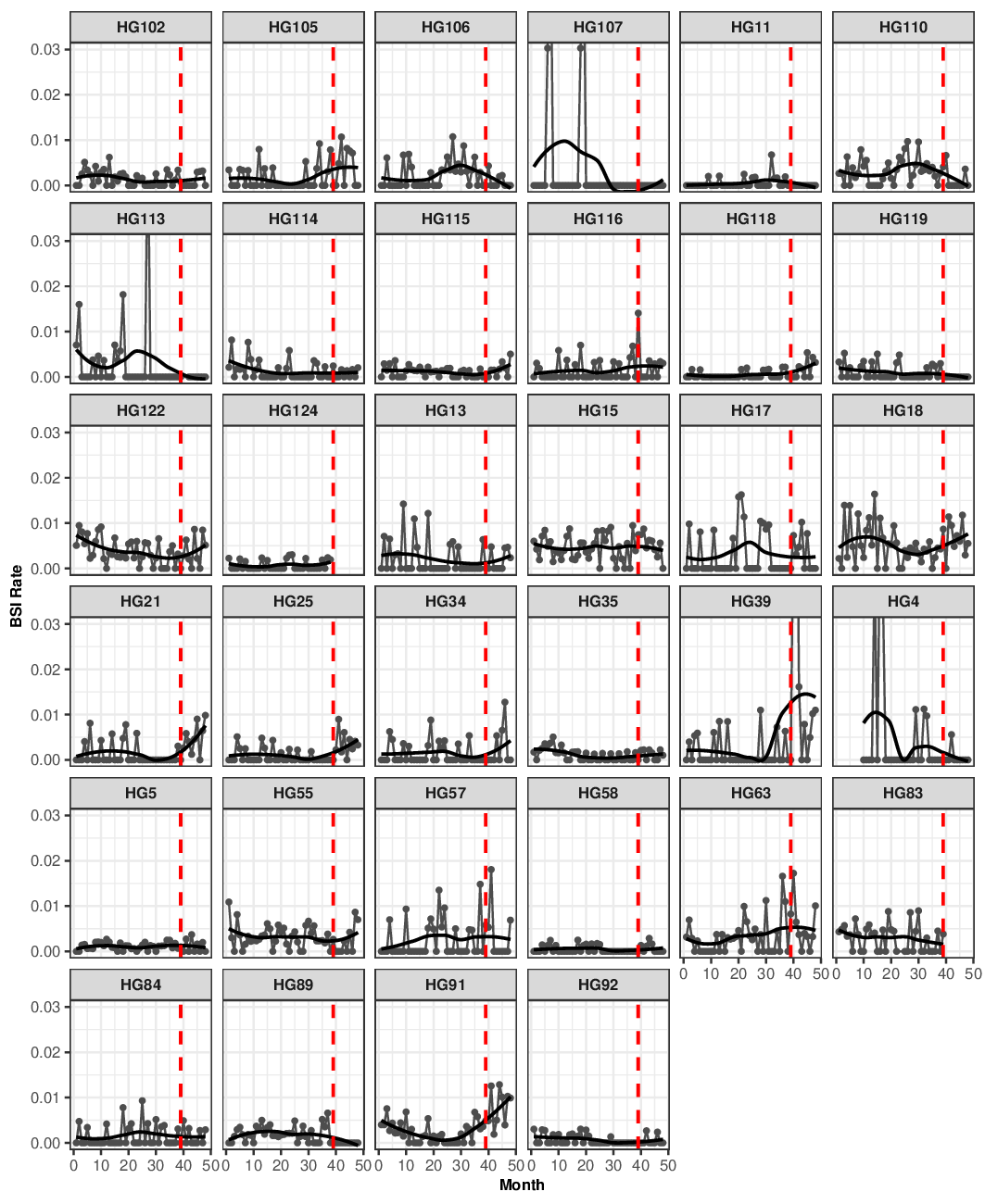}
    \caption{CR-BSI rate time series along with the LOESS curve by private hospitals. The COVID-19 pandemic is represented by the dashed line.}
     \label{hospital_time1}
\end{figure}

\begin{figure}
    \includegraphics[width=17cm,height=17cm]{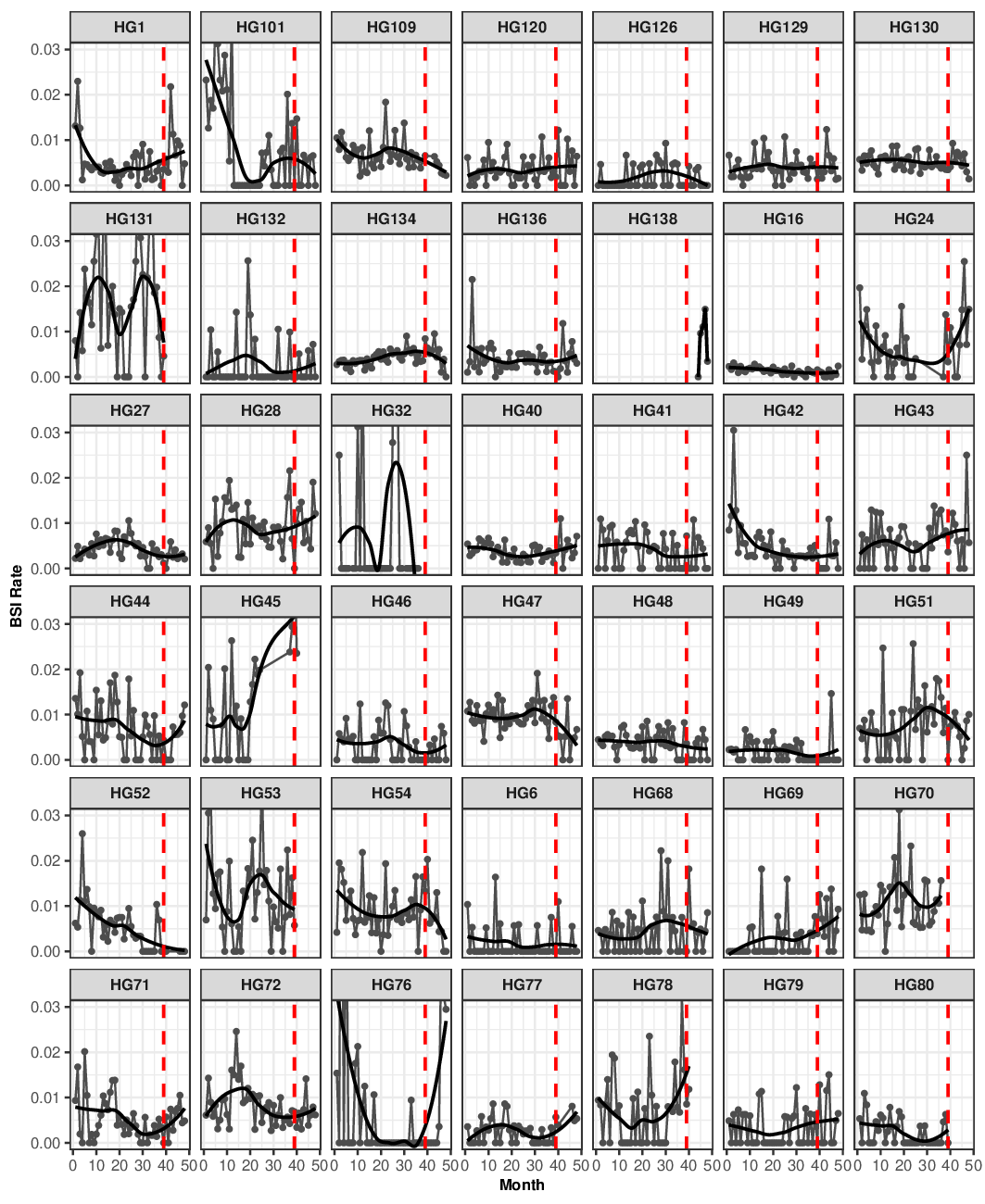}
    \caption{CR-BSI rate time series along with the LOESS curve by non-private hospitals. The COVID-19 pandemic is represented by the dashed line.}
     \label{hospital_time3}
\end{figure}

The Figures \ref{hospital_time1} and \ref{hospital_time3} show the temporal evolution of the CR-BSI rate with the LOESS curve for each hospital. The following can be observed: 1)for private hospitals, HG107, HG113, HG39, and HG4 show some unusual peaks, which may or may not represent data collection issues. In addition, hospitals HG124 and HG83 lack information on the CR-BSI rate after the pandemic; 2) for non-private hospitals, hospital HG76 also shows strange behavior, and hospital HG138 has information available only after the start of the COVID-19 pandemic. Hospitals HG32, HG131, HG45, HG70, HG78, HG80, and HG53 lack information after the onset of the COVID-19 pandemic.

Consequently, these mentioned hospitals were excluded from the dataset for modeling purposes. Therefore, based on EDA, the dataset for the analysis consists of $N=61$ hospitals, totaling $2,903$ observations.

\newpage
\section{Diagnostic plot - Section 5.2}
\begin{figure}[h]
     \centering
     \begin{subfigure}[b]{0.45\textwidth}
         \centering
         \includegraphics[width=\textwidth]{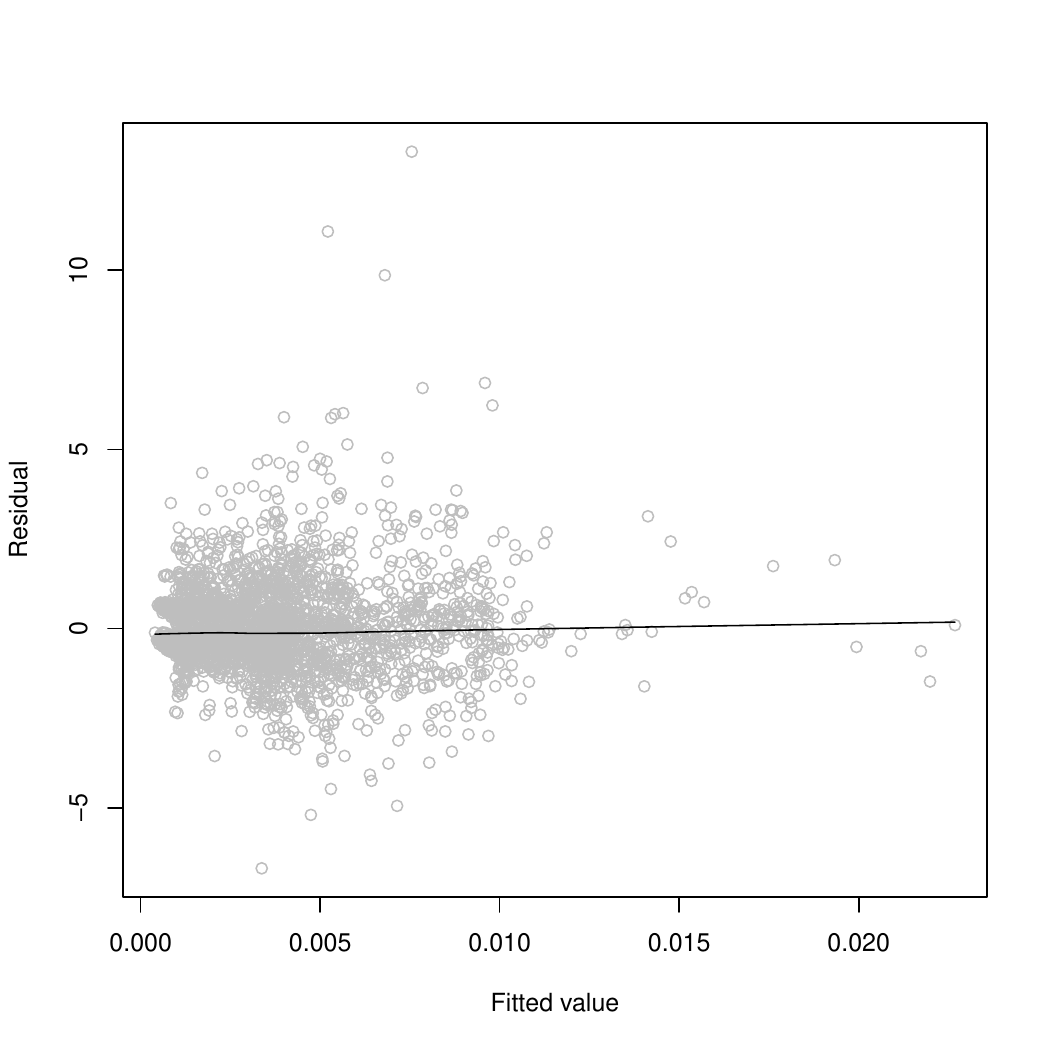}
         \caption{Residuals vs fitted values}
         \label{diagnostic_random1}
     \end{subfigure}
     \hfill
     \begin{subfigure}[b]{0.45\textwidth}
         \centering
         \includegraphics[width=\textwidth]{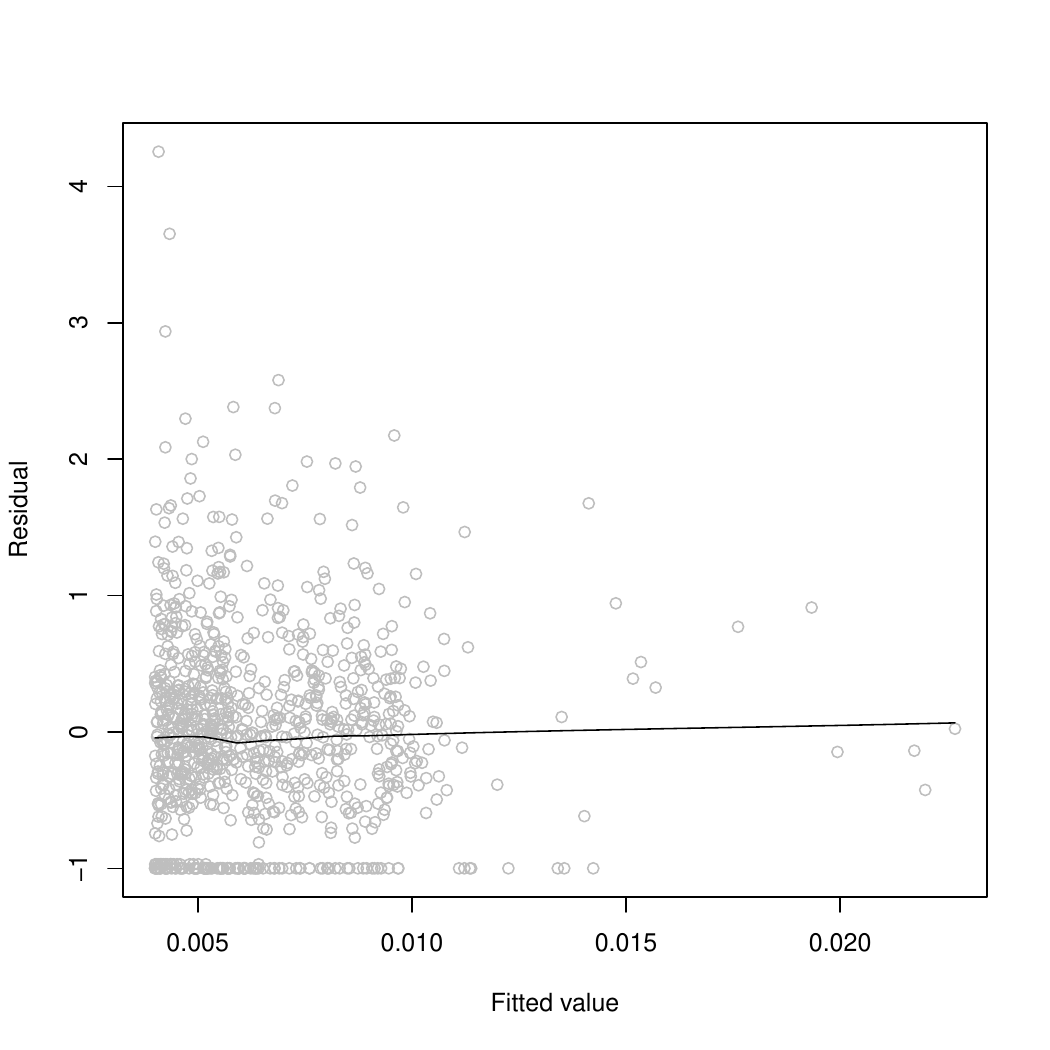}
         \caption{Working residuals vs fitted values}
         \label{diagnostic_random2}
     \end{subfigure}
     \vfill
     \begin{subfigure}[b]{0.48\textwidth}
         \centering
         \includegraphics[width=\textwidth]{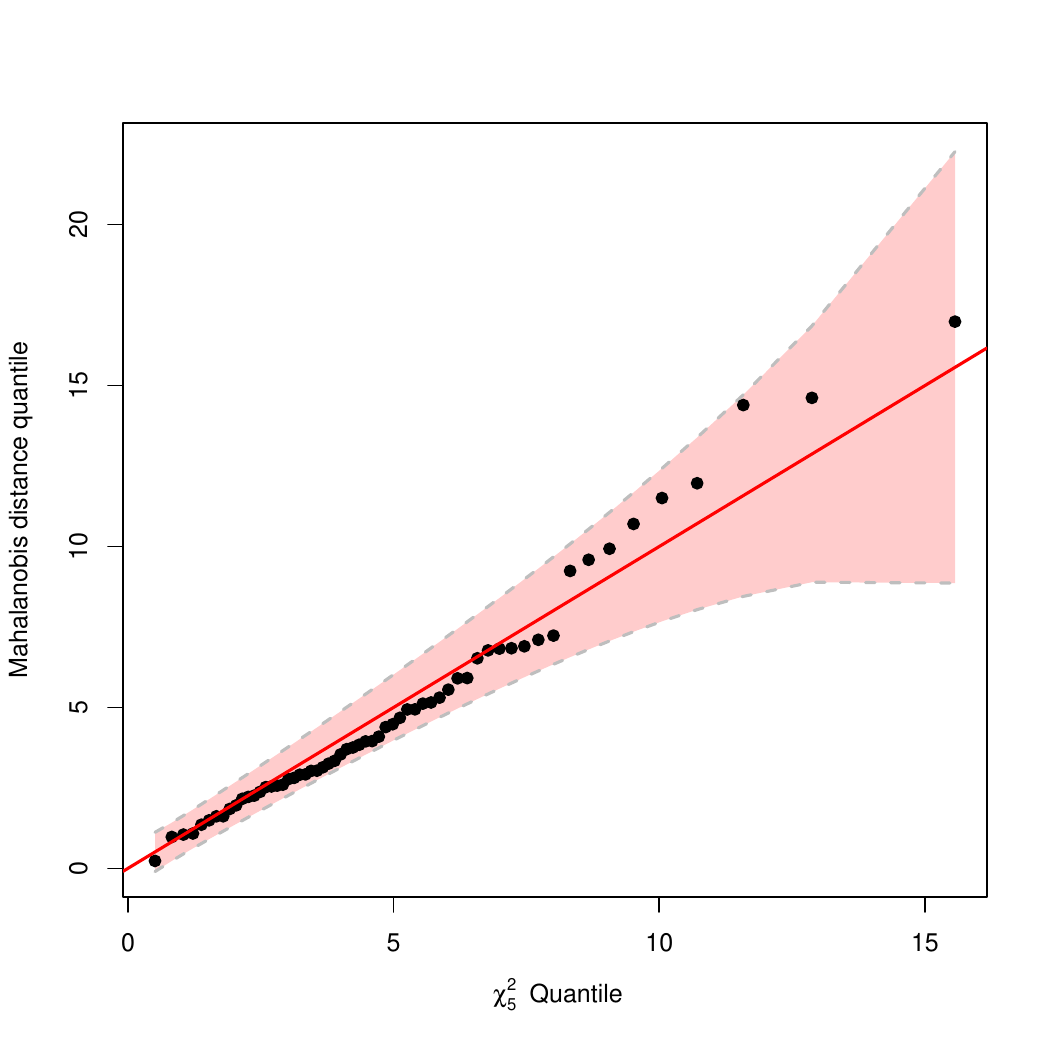}
         \caption{Mahalanobis distance QQ-plot for random effects}
         \label{diagnostic_random3}
     \end{subfigure}
        \caption{Diagnostic plot for the random changepoint case.}
        \label{diagnostic_random}
\end{figure}

Some diagnostic plots for the random changepoint case are showed in Figure \ref{diagnostic_random}.

